\documentclass[11pt,a4paper]{article}
\usepackage{myart}
\usepackage{amsmath}
\usepackage{graphicx}
\usepackage{amsfonts}
\usepackage{amssymb}

\begin{document}
\hfill\hbox{LPM-04-10}

\bigskip

\begin{center}
{\Large \textbf{On the Three-point Function in Minimal Liouville Gravity}}

\vspace{1.5cm}

{\large Contribution to the proceedings of the }

{\large International Workshop on Classical and Quantum Integrable Systems}

{\large \ Dubna, January 26--29, 2004.}

\vspace{1.5cm}

{\large Al.Zamolodchikov}\footnote{On leave of absence from Institute of
Theoretical and Experimental Physics, B.Cheremushkinskaya 25, 117259 Moscow, Russia.}

\vspace{0.2cm}

Laboratoire de Physique Math\'ematique\footnote{Laboratoire Associ\'e au CNRS URA-768}

Universit\'e Montpellier II

Pl.E.Bataillon, 34095 Montpellier, France
\end{center}

\vspace{1.0cm}

\textbf{Abstract}

The problem of the structure constants of the operator product expansions in
the minimal models of conformal field theory is revisited. We rederive these
previously known constants and present them in the form particularly useful in
the Liouville gravity applications. Analytic relation between our expression
and the structure constant in Liouville field theory is discussed. Finally we
present in general form the three- and two-point correlation numbers on the
sphere in the minimal Liouville gravity.

\section{Liouville gravity}

By gravity usually we imply a dynamic theory of the metric structure on
certain manifold. In the two dimensional case the latter is supposed to be a
two-dimensional surface $\Sigma$ (compact or non-compact) of certain topology
and equipped with a Riemann metric $g_{ab}(x)$. To avoid problems with moduli
in this paper I always imply that $\Sigma$ is a sphere. Also I consider the
euclidean version of the gravity, i.e., $g_{ab}$ is always non-degenerate and
has positive signature. In the path integral approach the problem is reduced
to the evaluation of a functional integral over all Riemann metric $D[g]$
modulo the diffeomorphism equivalent $g_{ab}(x)$. E.g., the gravitational
partition function of a sphere is formally written as
\begin{equation}
Z=\int D[g]\exp\left(  -A_{\text{eff}}[g]\right) \label{Zg}%
\end{equation}
Here $A_{\text{eff}}[g]$ is supposed to be an effective action induced by some
generally covariant ``matter'' field theory living on the surface. General
covariance ensures that $A_{\text{eff}}[g]$ is invariant under diffeomorphisms.

In general case of massive matter $A_{\text{eff}}[g]$ is a non-local and quite
complicated functional of the metric. The problem (\ref{Zg}) appears very
complicated. There is a drastic simplification however, if all the matter
inducing $A_{\text{eff}}[g]$ is ``critical'', i.e., described by a conformal
field theory (CFT). In this case the form of $A_{\text{eff}}[g]$ is very
universal and simple, being called the Liouville action. This fact was first
discovered by A.Polyakov in 1981 \cite{Polyakov} by direct computations with
free fields. In general CFT this statement simply follows from the form of
conformal anomaly. Moreover, conceptually this form of effective action can be
taken (with few additional assumptions) as the very definition of CFT.

Due to the diffeomorphism invariance of $A_{\text{eff}}[g]$ there is a gauge
fixing problem in (\ref{Zg}). One of the most convenient gauge choices is the
conformal gauge where the coordinates on $\Sigma$ are (locally) chosen in the
way that (this is always possible in two dimensions)
\begin{equation}
g_{ab}(x)=e^{2b\phi(x)}\delta_{ab}\label{conf}%
\end{equation}
The scale factor here is described by a quantum field $\phi(x)$ called the
Liouville field (see below for the definition of the parameter $b$). One can
also fix the gauge in a covariant way choosing an arbitrary metric
$g_{ab}^{(0)}(x)$ as the reference one and requiring
\begin{equation}
g_{ab}(x)=e^{2b\phi(x)}g_{ab}^{(0)}(x)\label{background}%
\end{equation}
In the latter approach the Liouville field $\phi$ is a usual scalar under
coordinate transformations. Gauge (\ref{conf}) then implies a particular
choice of coordinate system and setting $g_{ab}^{(0)}(x)=\delta_{ab}$ in these coordinates.

As usual, the gauge fixing introduces the Faddeev-Popov determinant. In our
case it can be described by the $BC$ system of spin $(2,-1)$%
\begin{equation}
A_{\text{gh}}=\frac1\pi\int(C\bar\partial B+\bar C\partial\bar B)d^{2}%
x\label{Agh}%
\end{equation}
This is again a conformal field theory with central charge $c_{\text{gh}}=-26
$ and therefore the gauge fixing determinant is reduced again to the Liouville
action. This fact was also observed in \cite{Polyakov}. The gravitational
partition function reduces to
\begin{equation}
Z_{\text{g}}=\int D[\phi]\exp\left(  -A_{\text{L}}[\phi]\right) \label{ZLG}%
\end{equation}
where $A_{\text{L}}[\phi]$ is the Liouville action induced by the matter
fields and ghosts.

There is certain problem with the integration measure $D[\phi]$ over the
Liouville field configurations \cite{Polyakov}. Complete definition of the
path integral (\ref{ZLG}) requires ultraviolet cutoff, which, from the
physical point of view, must depend itself on the scale factor $\exp(2b\phi)$.
This means that the integration measure differs from the ordinary (linear)
integration measure where the cutoff is defined with respect to certain fixed
metric. The direct evaluation of (\ref{ZLG}) with this non-linear measure
turns out quite difficult both technically and conceptually. However, in
refs.\cite{DK, FD} it was suggested that the effect of this complicated
non-linear measure can be reduced to certain finite renormalization of the
parameters. This means that in (\ref{ZLG}) ordinary linear measure (with
respect to fixed reference metric) can be consistently used once the
parameters in $A_{\text{L}}[\phi]$ are chosen properly. Then the renormalized
parameters can be determined from the consistency conditions. This assumption
is not in fact well justified theoretically. The only serious support might
come from actual calculations in this framework and comparison of the results
with other known facts in 2D quantum gravity. Among them are the results of
the discrete, or matrix model approach (see e.g. the reviews \cite{matrix} and
references therein) and the field-theoretic calculations in different gauges,
i.e., in the light-cone gauge \cite{KPZ} (called sometimes the Polyakov gauge)
where the above problem of non-linear measure is less important.

After this procedure done the renormalized Liouville action comes out as
\begin{equation}
A_{\text{L}}[\phi]=\int\left(  \frac1{4\pi}\left(  \partial_{a}\phi\right)
^{2}+\mu e^{2b\phi}\right)  d^{2}x\label{AL}%
\end{equation}
where $\mu$ is yet another parameter called the cosmological constant. Action
(\ref{AL}) describes again a conformal field theory with central charge
\begin{equation}
c_{\text{L}}=1+6Q^{2}\label{cL}%
\end{equation}
where the ``background charge'' $Q$ is related to the parameter $b$ as
\begin{equation}
Q=b+b^{-1}\label{Q}%
\end{equation}
The consistency condition of David and Distler-Kawai, mentioned above, fixes
the parameter $b$ from the condition that the total central charge of the
joint CFT of matter, Liouville and ghosts
\begin{equation}
A_{\text{g}}=A_{\text{CFT}}+A_{\text{L}}+A_{\text{gh}}\label{Ag}%
\end{equation}
vanishes (which means in fact the independence of physical observables on the
choice of reference metric in (\ref{background}))
\begin{equation}
c_{\text{M}}+c_{\text{L}}=26\label{cbalance}%
\end{equation}
Here $c_{\text{M}}$ is the central charge of the matter CFT and $A_{\text{CFT}%
}$ stands for its formal action.

The action (\ref{Ag}) implies that in critical gravity the three field
theories (conformal matter, Liouville field $\phi$ and ghosts) are formally
decoupled, interacting only through the conformal anomaly. In the correlation
functions any matter primary field $\Phi$ of dimension $\Delta$ should be
consistently ``dressed'' by an appropriate Liouville exponential field
$\exp\left(  2a\phi\right)  $ to form a composite field of dimension $(1,1)$
\begin{equation}
O=\Phi e^{2a\phi}\label{composite}%
\end{equation}
In Liouville field theory (\ref{AL}) the exponential $\exp\left(
2a\phi\right)  $ has dimension
\begin{equation}
\Delta_{a}=a(Q-a)\label{DL}%
\end{equation}
so that the dressing parameter $a$ is determined from the condition
\begin{equation}
\Delta+\Delta_{a}=1\label{Dbalance}%
\end{equation}

The correlation functions of the dressed operators are thus decoupled (before
integration over moduli) to a product of Liouville and matter correlation
functions
\begin{equation}
\left\langle O_{1}(x_{1})\ldots O_{n}(x_{n})\right\rangle _{\text{LG}%
}=\left\langle \Phi_{1}(x_{1})\ldots\Phi_{n}(x_{n})\right\rangle _{\text{CFT}%
}\left\langle e^{2a_{1}\phi}(x_{1})\ldots e^{2a_{n}\phi}(x_{n})\right\rangle
_{\text{L}}\label{LGn}%
\end{equation}
where $\left\langle \ldots\right\rangle _{\text{LG}}$, $\left\langle
\ldots\right\rangle _{\text{CFT}}$ and $\left\langle \ldots\right\rangle
_{\text{L}}$ stand for the correlation functions in Liouville gravity, matter
CFT and Liouville theory respectively.

It is well known that on the sphere, due to the existence of excessive
conformal Killing vector fields, the conformal gauge needs some further
fixing. This can be achieved by decorating arbitrary three fields in the
correlation function (\ref{LGn}) by ghost multipliers $C\bar C$ which give the
product a total dimension $(0,0)$. The leftover insertions have dimensions
$(1,1)$ and can be integrated over their coordinates. The resulting
correlation function
\begin{align}
\  &  \left\langle O_{1}\ldots O_{n}\right\rangle _{\text{G}}=\label{Ogn}\\
&  \ \int\left\langle C\bar C(x_{1})\ldots CC(x_{3})\right\rangle _{\text{gh}%
}\left\langle \Phi_{1}(x_{1})\ldots\Phi_{n}(x_{n})\right\rangle _{\text{CFT}%
}\left\langle e^{2a_{1}\phi}(x_{1})\ldots e^{2a_{n}\phi}(x_{n})\right\rangle
_{\text{L}}d^{2}x_{4}\ldots d^{2}x_{n}\nonumber
\end{align}
depends no more on the coordinates (and therefore on the choice of the gauge)
and is better named the correlation number. To indicate this we'll use the
notation $\left\langle \ldots\right\rangle _{\text{G}}$ for them. These gauge
invariant correlation numbers are among the main objects of interest in
two-dimensional gravity. For example, suppose a more complicated gravity is
addressed, where the matter on the surface is not purely critical but
described as a certain perturbation of a conformal field theory
\begin{equation}
A_{\text{M}}=A_{\text{CFT}}+\lambda\int\Phi(x)d^{2}x\label{Alambda}%
\end{equation}
Here $\Phi$ a relevant primary matter field (of dimension $\Delta$) and
$\lambda$ is the corresponding coupling constant. In the gravity environment
$\Phi$ is dressed by $\exp\left(  2a\phi\right)  $ (where $a$ is determined by
(\ref{DL}) and (\ref{Dbalance})). The perturbative development in $\lambda$%
\begin{equation}
Z(\lambda,\mu)=Z(0,\mu)\sum_{n=0}^{\infty}\frac{(-\lambda)^{n}}{n!}%
a_{n}\label{Zlambda}%
\end{equation}
where
\begin{align}
a_{0}  &  =1\label{an}\\
a_{n}  &  =\int\left\langle C\bar C(x_{1})\ldots CC(x_{3})\right\rangle
_{\text{gh}}\left\langle \Phi(x_{1})\ldots\Phi(x_{n})\right\rangle
_{\text{CFT}}\left\langle e^{2a\phi}(x_{1})\ldots e^{2a\phi}(x_{n}%
)\right\rangle _{\text{L}}d^{2}x_{4}\ldots d^{2}x_{n}\;;\;\;\;n\geq3\nonumber
\end{align}
(the cases $n=1$ and $n=2$ are somewhat special and will be commented
separately) is thus expressed in terms of multipoint correlation numbers
(\ref{Ogn}).

In this article we're interested in a special case of 2D induced critical
gravity, the minimal gravity (MG). This term means that the matter content
consists of a single minimal model of CFT \cite{BPZ} $\mathcal{M}%
_{p,p^{\prime}}$ with $(p,p^{\prime})$ a pair of coprime entire numbers. It is
strongly suggested by the results of the discrete approach to 2D gravity
\cite{matrix} that the minimal gravity is exactly solvable. E.g., the scaling
functions like (\ref{Zlambda}) and therefore the correlation numbers
(\ref{Ogn}) come out in the matrix model framework explicitly. For the
Liouville gravity it is still a challenge to reveal its potential in
reproducing the exact results of discrete methods.

This note is devoted to the simplest correlation number in MG, the three-point
one. In this case in eq.(\ref{Ogn}) there are no integrations to perform and
our task is simply to multiply the Liouville three-point function by the one
in $\mathcal{M}_{p,p^{\prime}}.$ This simple job is performed in the
subsequent sections with some comments about analytic relation between minimal
models and Liouville field theory.

Before turning to this content, let me mention that many of the results seem
to be related to those of the forthcoming paper ``Non-rational 2D quantum
gravity: ground ring versus loop gas'' by I.Kostov and V.Petkova \cite{KP}.

\section{Liouville three-point function}

The three-point function of exponential fields $\exp(2a\phi)$%
\begin{equation}
\left\langle e^{2a_{1}\phi}(x_{1})e^{2a_{2}\phi}(x_{2})e^{2a_{3}\phi}%
(x_{3})\right\rangle _{\text{L}}=\frac{C_{\text{L}}(a_{1},a_{2},a_{3}%
)}{\left(  x_{12}\bar x_{12}\right)  ^{\Delta_{1}+\Delta_{2}-\Delta_{3}%
}\left(  x_{23}\bar x_{23}\right)  ^{\Delta_{2}+\Delta_{3}-\Delta_{1}}\left(
x_{31}\bar x_{31}\right)  ^{\Delta_{3}+\Delta_{1}-\Delta_{2}}}\label{L3}%
\end{equation}
in Liouville field theory has been discovered by Dorn and Otto \cite{DO} in
1992. The coordinate dependence of (\ref{L3}) involves the dimensions
$\Delta_{i}=\Delta_{a_{i}}$ of the exponential fields given by eq.(\ref{DL}).
The dependence is standard and therefore we'll omit this multiplier and call
the factor $C_{\text{L}}(a_{1},a_{2},a_{3})$ the three-point function. In the
notations of ref. \cite{ZZ1} it reads explicitly
\begin{align}
C_{\text{L}}(a_{1},a_{2},a_{3})  &  =\left(  \pi\mu\gamma(b^{2})b^{2-2b^{2}%
}\right)  ^{(Q-a_{1}-a_{2}-a_{3})/b}\times\label{CL}\\
&  \ \frac{\Upsilon(b)\Upsilon(2a_{1})\Upsilon(2a_{2})\Upsilon(2a_{3}%
)}{\Upsilon(a_{1}+a_{2}+a_{3}-Q)\Upsilon(a_{1}+a_{2}-a_{3})\Upsilon
(a_{2}+a_{3}-a_{1})\Upsilon(a_{3}+a_{1}-a_{2})}\nonumber
\end{align}
Here $\Upsilon(x)=\Upsilon_{b}(x)$ is a special function related to the Barnes
double gamma function \cite{Barnes} (see \cite{ZZ1} for the precise
definitions and properties). It should be noted at this point that (\ref{CL})
is unnormalized correlation function. To use it in the developments like
(\ref{Zlambda}) one has to divide it by the Liouville partition function of
the sphere. Below we'll comment on this point.

Later in 1995 expression (\ref{CL}) has been rederived by J.Teschner
\cite{Teschner1} in a more systematic way by means of the conformal bootstrap
technique. It turns out that self-consistency (bootstrap) requires
$C_{\text{L}}(a_{1},a_{2},a_{3})$ to satisfy the following functional
relations
\begin{align}
&  \frac{C_{\text{L}}(a_{1}+b,a_{2},a_{3})}{C_{\text{L}}(a_{1},a_{2},a_{3}%
)}=-\frac{\gamma(-b^{2})}{\pi\mu}\times\nonumber\\
&  \frac{\gamma(2a_{1}b+b^{2})\gamma(2a_{1}b)\gamma(b(a_{2}+a_{3}-a_{1}%
)-b^{2})}{\gamma(b(a_{1}+a_{2}+a_{3})-1-b^{2})\gamma(b(a_{1}+a_{2}%
-a_{3}))\gamma(b(a_{1}+a_{3}-a_{2}))}\label{CLshift}\\
&  \frac{C_{\text{L}}(a_{1}+b^{-1},a_{2},a_{3})}{C_{\text{L}}(a_{1}%
,a_{2},a_{3})}=-\frac{\gamma(-b^{-2})}{\pi\tilde\mu}\times\nonumber\\
&  \frac{\gamma(2a_{1}b^{-1}+b^{-2})\gamma(2a_{1}b^{-1})\gamma(b^{-1}%
(a_{2}+a_{3}-a_{1})-b^{-2})}{\gamma(b^{-1}(a_{1}+a_{2}+a_{3})-1-b^{-2}%
)\gamma(b^{-1}(a_{1}+a_{2}-a_{3}))\gamma(b^{-1}(a_{1}+a_{3}-a_{2}))}\nonumber
\end{align}
Here $\tilde\mu$ is the ``dual cosmological constant'' related to $\mu$ as
\cite{ZZ1}
\begin{equation}
\left(  \tilde\mu\gamma(b^{-2})\right)  ^{b}=\left(  \mu\gamma(b^{2})\right)
^{1/b}\label{mudual}%
\end{equation}
In the general case $b$ and $b^{-1}$ are incommensurable and (\ref{CL}) is the
unique solution to this system.

Expression (\ref{CL}) implies special normalization of the Liouville
exponential fields. It is fixed by the two-point function
\begin{equation}
\left\langle e^{2a\phi}(x)e^{2a\phi}(0)\right\rangle _{\text{L}}%
=\frac{D_{\text{L}}(a)}{\left(  x\bar x\right)  ^{2\Delta_{a}}}\label{L2}%
\end{equation}
with
\begin{equation}
D_{\text{L}}(a)=\frac{\left(  \pi\mu\gamma(b^{2})\right)  ^{(Q-2a)/b}}{b^{2}%
}\frac{\gamma(2ab-b^{2})}{\gamma(2-2ab^{-1}+b^{-2})}\label{Da}%
\end{equation}
This is again a non-normalized expression. How to use it in the two-point
correlation number will be also discussed below. This quantity enters the
diagonal metric in the space of exponential fields, so that the structure
constant reads
\begin{equation}
C_{a_{1}a_{2}}^{\text{(L)}a_{3}}=D^{-1}(a_{3})C_{\text{L}}(a_{1},a_{2}%
,a_{3})=C_{\text{L}}(a_{1},a_{2},Q-a_{3})\label{Lstruct}%
\end{equation}

For subsequent references let me quote explicitly the special structure
constants
\begin{align}
C_{+}^{\text{(L)}}(a)  &  =\tilde C_{+}^{\text{(L)}}(a)=1\nonumber\\
C_{-}^{\text{(L)}}(a)  &  =-\frac{\pi\mu}{\gamma(-b^{2})}\frac{\gamma
(2ab-b^{2}-1)}{\gamma(2ab)}\label{CLpm}\\
\tilde C_{-}^{\text{(L)}}(a)  &  =-\frac{\pi\tilde\mu}{\gamma(-b^{-2})}%
\frac{\gamma(2ab^{-1}-b^{-2}-1)}{\gamma(2ab^{-1})}\nonumber
\end{align}
appearing in the truncated operator product expansions of the level two
degenerate exponentials $\exp(-b\phi)$ and $\exp(-b^{-1}\phi)$
\begin{align}
e^{-b\phi}e^{2a\phi}  &  =C_{+}^{\text{(L)}}(a)\left[  e^{(2a-b)\phi}\right]
+C_{-}^{\text{(L)}}(a)\left[  e^{(2a+b)\phi}\right] \label{V12}\\
e^{-b^{-1}\phi}e^{2a\phi}  &  =\tilde C_{+}^{\text{(L)}}(a)\left[
e^{(2a-b^{-1})\phi}\right]  +\tilde C_{-}^{\text{(L)}}(a)\left[
e^{(2a+b^{-1})\phi}\right] \nonumber
\end{align}
Here and below in the r.h.s of operator product expansions we conventionally
omit the standard $x$-dependent multipliers and denote by $[\Phi]$ the
contribution of a primary field $\Phi$ and all its conformal descendants.

\section{Generalized minimal models}

CFT minimal model $\mathcal{M}_{p,p^{\prime}}$ is characterized by the central
charge
\begin{equation}
c_{\text{M}}=1-6(\beta^{-1}-\beta)^{2}\label{cM}%
\end{equation}
where
\begin{equation}
\beta=\sqrt{p/p^{\prime}}\label{betapp}%
\end{equation}
We'll imply below that $p<p^{\prime}$ so that $\beta<1$. It will be also
convenient to introduce the parameter
\begin{equation}
q=\beta^{-1}-\beta\label{q}%
\end{equation}
Thus \cite{BPZ}
\begin{equation}
c_{p/p^{\prime}}=1-6q^{2}=1-\frac{6(p-p^{\prime})^{2}}{pp^{\prime}}\label{cpp}%
\end{equation}
Principal (unitary) series corresponds to $p^{\prime}=p+1$.

In the minimal models parameter $\beta^{2}$ is supposed to be a rational
number. However, in the present study we'd like to look for as much
analyticity as possible. Therefore we'll try to consider $\beta$ as a
continuous real (and later even complex) parameter as far as it will rise no
apparent problems. The corresponding continuous family of (formal) CFT models
is called the generalized minimal model (GMM) and denoted $\mathcal{M}%
_{\beta^{2}}$.

The primary operator content of minimal model is given by the set of
degenerate fields $\Phi_{m,n}$, where $(m,n)$ is a pair of entire numbers.
Their dimensions are given by the Kac formula \cite{Kac}
\begin{equation}
\Delta_{m,n}=\frac{(m\beta^{-1}-n\beta)^{2}-(\beta^{-1}-\beta)^{2}}%
4=\alpha_{m,n}(\alpha_{m,n}-q)\label{Dmn}%
\end{equation}
In the last parametrization in terms of $\alpha_{m,n}$ there are two
possibilities to choose $\alpha_{m,n}$ which differ in the replacement
$\alpha_{m,n}\leftrightarrow q-\alpha_{m,n}$. To be definite we take below
\begin{equation}
\alpha_{m,n}=\frac{(n-1)\beta}2-\frac{(m-1)\beta^{-1}}2\label{amn}%
\end{equation}

The representations of the Virasoro algebra with central charge (\ref{cM}) and
dimensions (\ref{Dmn}) are all degenerate with a null-vector at level $mn $.
In minimal models all these null-vectors are supposed to vanish. This is one
of the basic requirements in the construction of minimal models \cite{BPZ}. It
results in certain linear differential equations for the degenerate primary
fields. For example, the simplest non-trivial fields $\Phi_{1,2}$ and
$\Phi_{2,1}$ satisfy
\begin{align}
\left(  \partial^{2}-\beta^{2}T\right)  \Phi_{1,2}  &  =0\label{diff12}\\
\left(  \partial^{2}\Phi_{21}-\beta^{-2}T\right)  \Phi_{2,1}  &  =0\nonumber
\end{align}
and similar ``left'' equations with $\partial$ replaced by $\bar\partial$ and
the right stress-tensor component $T=T_{zz}$ by the left one $\bar T=T_{\bar
z\bar z}$ (see \cite{BPZ} for more detailed analysis). In order, these
differential equations restrict the form of the operator product expansions of
the degenerate fields $\Phi_{m,n}$ to a very special form \cite{BPZ}
\begin{equation}
\Phi_{m_{1},n_{1}}\Phi_{m_{2},n_{2}}=\sum_{(m_{3},n_{3})}C_{(m_{1}%
,n_{1})(m_{2},n_{2})}^{(m_{3},n_{3})}\left[  \Phi_{m_{3},n_{3}}\right]
\label{minOPE}%
\end{equation}

The structure constants $C_{(m_{1},n_{1})(m_{2},n_{2})}^{(m_{3},n_{3})}$
depend of course on the normalization of the fields $\Phi_{m,n}$. In this
paper we adopt the standard CFT normalization through the two-point functions
\begin{equation}
\left\langle \Phi_{m,n}(x)\Phi_{m,n}(0)\right\rangle _{\text{CFT}}%
=\frac1{\left(  x\bar x\right)  ^{2\Delta_{m,n}}}\label{CFTmn}%
\end{equation}
In this normalization the structure constants coincide with the three-point
functions
\begin{equation}
C_{(m_{1},n_{1})(m_{2},n_{2})}^{(m_{3},n_{3})}=C_{(m_{1},n_{1})(m_{2}%
,n_{2})(m_{3},n_{3})}=\left\langle \Phi_{m_{1},n_{1}}\Phi_{m_{2},n_{2}}%
\Phi_{m_{3},n_{3}}\right\rangle _{\text{CFT}}\label{CFT3point}%
\end{equation}
(again we omit here the standard $x$-dependent factor).

The structure constants in GMM are further restricted by the fusion rules,
which also follow from the vanishing of the null-vectors. The fusion rules
restrain separately the possible triples of indices $(m_{1},m_{2},m_{3})$ and
$(n_{1},n_{2},n_{3})$ in the structure constants. In fact they are equivalent
to the fusion algebra of the regular representations of $SL(2)$. Thus the GMM
structure constants are proportional to the fusion algebra structure
constants
\begin{equation}
C_{(m_{1},n_{1})(m_{2},n_{2})(m_{3},n_{3})}\sim f_{m_{1},m_{2},m_{3}}%
f_{n_{1},n_{2},n_{3}}\label{ff}%
\end{equation}
where
\begin{equation}
f_{m_{1},m_{2},m_{3}}=\left\{
\begin{array}
[c]{c}%
\;\;\;\;\;\;\;\;\;\;1\;\;\;\text{if\ \ }\left\{
\begin{array}
[c]{c}%
m_{1}+m_{2}-m_{3}\geq0\\
m_{2}+m_{3}-m_{1}\geq0\\
m_{3}+m_{1}-m_{2}\geq0
\end{array}
\right. \\
0\;\;\;\;\;\;\;\;\;\;\;\;\;\;\;\;\;\text{otherwise}%
\end{array}
\right. \label{fmmm}%
\end{equation}

The GMM structure constants $C_{(m_{1},n_{1})(m_{2},n_{2})(m_{3},n_{3})}$ has
been computed explicitly as certain products of gamma functions by Dotsenko
and Fateev \cite{DF}. In the next section we'll try to rederive their result
in a slightly different form which would conform more with our search of
analyticity. To do that we introduce instead of the discrete set of degenerate
operators $\Phi_{m,n}$ a formal family of fields $\Phi_{\alpha} $
parameterized by continuous parameter $\alpha$. Their dimensions are
\begin{equation}
\Delta_{\alpha}=\alpha(\alpha-q)\label{Dalpha}%
\end{equation}
As $\Delta_{\alpha}=\Delta_{q-\alpha}$ it seems natural to ``fold'' the family
of formal fields by the equivalence
\begin{equation}
\Phi_{\alpha}=\Phi_{q-\alpha}\label{q-a}%
\end{equation}
Then we'll try to use the bootstrap technique to determine a formal
three-point function (again the coordinate dependence is omitted)
\begin{equation}
C_{\text{M}}(\alpha_{1},\alpha_{2},\alpha_{3})=\left\langle \Phi_{\alpha_{1}%
}\Phi_{\alpha_{2}}\Phi_{\alpha_{3}}\right\rangle _{\text{M}}\label{CMPhi}%
\end{equation}
as a continuous function of its three parameters, such that when specialized
to particular values (\ref{amn}) $\alpha_{i}=\alpha_{m_{i},n_{i}}$ it would
reproduce the known Dotsenko-Fateev expressions.

Before turning to this task, three remarks are in order

1. In a genuine minimal model $\mathcal{M}_{p,p^{\prime}}$ there is a finite
subset of degenerate primaries $\Phi_{m,n}$ with $1\leq m\leq p-1$ and $1\leq
n\leq p^{\prime}-1$, closed with respect to the operator product algebra (the
fields $\Phi_{m,n}$ and $\Phi_{p-m,p^{\prime}-n}$ being identified, so that
the finite subset counts in total $(p-1)(p^{\prime}-1)/2 $ fields) \cite{BPZ}.
Therefore in the genuine minimal models it is possible to consistently reduce
the operator content $\mathcal{M}_{\beta^{2}}$ to this finite subset of
primary fields thus arriving at a rational conformal field theory. In GMM with
irrational $\beta^{2}$ such reduction is impossible and one is forced to
consider the whole set of $\Phi_{m,n}$ with $(m,n)$ arbitrary positive
integers. It remains a question if such infinite algebra is consistent with
general requirements of quantum field theory. In particular, the construction
of a modular invariant partition function of GMM obviously encounters severe
problems. In the present study we prefer to stay quite formal and forget for a
while about these important questions.

2. It will be seen below, that the continuous function $C_{\text{M}}%
(\alpha_{1},\alpha_{2},\alpha_{3})$ (which will be constructed below) when
specialized to the degenerate values $\alpha_{i}=\alpha_{m_{i},n_{i}}$ doesn't
always vanish automatically if the fusion rules (\ref{fmmm}) are violated.
Sometimes it gives certain finite numbers whose interpretation remains
mysterious for me for the time being. Hence, to obtain the correct set of GMM
structure constants we have to take the fusion rules into account separately.
The relation between $C_{(m_{1},n_{1})(m_{2},n_{2})(m_{3},n_{3})} $ and
$C_{\text{M}}(\alpha_{1},\alpha_{2},\alpha_{3})$ then appears as
\begin{equation}
C_{(m_{1},n_{1})(m_{2},n_{2})(m_{3},n_{3})}=f_{m_{1},m_{2},m_{2}}%
f_{n_{1},n_{2},n_{2}}C_{\text{M}}(\alpha_{m_{1},n_{1}},\alpha_{m_{2},n_{2}%
},\alpha_{m_{3},n_{3}})\label{CmmmC}%
\end{equation}
Apparently, this decoration is not along with our search of maximal
analyticity. However, for the moment I don't see any way to avoid this manual
way to impose the fusion rules.

3. It is easy to see that the expressions for the central charges (\ref{cL})
and (\ref{cM}) as well as for the primary field dimensions (\ref{DL}) and
(\ref{Dalpha}) in the Liouville field theory and GMM are simply related by the
analytic continuation of the parameters
\begin{align}
b  &  =-i\beta\nonumber\\
a  &  =i\alpha\label{btobeta}\\
Q  &  =iq\nonumber
\end{align}
Together with more analytic relations to be observed below, this might give
rise to an idea that GMM is simply an analytic continuation of the Liouville
field theory for pure imaginary values of the parameter $b$ (or vise versa).
We'll see shortly that this guess is incompatible with our continuous
approach. In particular $C_{\text{M}}(\alpha_{1},\alpha_{2},\alpha_{3})$ is
not an analytic continuation of $C_{\text{L}}(a_{1},a_{2},a_{3})$.

\section{Conformal bootstrap in GMM}

Even for the formal continuous fields $\Phi_{\alpha}$ the null-vector
decoupling drastically restricts the form of operator product expansions. In
particular, eqs.(\ref{diff12}) imply that
\begin{align}
\Phi_{1,2}\Phi_{\alpha}  &  =C_{+}^{\text{(M)}}(\alpha)\left[  \Phi
_{\alpha+\beta/2}\right]  +C_{-}^{\text{(M)}}(\alpha)\left[  \Phi
_{\alpha-\beta/2}\right] \label{Phi12Phia}\\
\Phi_{2,1}\Phi_{\alpha}  &  =\tilde C_{+}^{\text{(M)}}(\alpha)\left[
\Phi_{\alpha-\beta^{-1}/2}\right]  +\tilde C_{-}^{\text{(M)}}(\alpha)\left[
\Phi_{\alpha+\beta/2}\right] \nonumber
\end{align}
Here $C_{\pm}^{\text{(M)}}(\alpha)=C_{\text{M}}(\alpha_{1,2},\alpha,\alpha
\pm\beta/2)$ and $\tilde C_{\pm}^{\text{(M)}}(\alpha)=C_{\text{M}}%
(\alpha_{2,1},\alpha,\alpha\mp\beta/2)$ are special structure constants,
related to the continuous function $C_{\text{M}}(\alpha_{1},\alpha_{2}%
,\alpha_{3})$ through appropriate specializations.

Below we'll follow the standard technique of finite-dimensional bootstrap,
developed in \cite{BPZ} and many subsequent works. The calculations follow
almost literally the Liouville related development by J.Teschner
\cite{Teschner1}. They are recapitulated in the Appendix A and result in two
functional relations for $C_{\text{M}}(\alpha_{1},\alpha_{2},\alpha_{3})$
\begin{align}
\  &  \frac{C_{+}^{\text{(M)}}(\alpha_{1})C_{\text{M}}(\alpha_{1}%
+\beta/2,\alpha_{2},\alpha_{3})}{C_{-}^{\text{(M)}}(\alpha_{1})C_{\text{M}%
}(\alpha_{1}-\beta/2,\alpha_{2},\alpha_{3})}=\label{F1}\\
\  &  =-\dfrac{(1-\beta^{2}-2\alpha_{1}\beta)^{2}\gamma(\beta^{2}%
/2+(\alpha_{3}+\alpha_{1}-\alpha_{2})\beta)\gamma(\beta^{2}/2+(\alpha
_{1}+\alpha_{2}-\alpha_{3})\beta)}{\gamma^{2}(\beta^{2}+2\alpha_{1}%
\beta)\gamma(\beta^{2}/2+(\alpha_{2}+\alpha_{3}-\alpha_{1})\beta
)\gamma(2-3\beta^{2}/2-(\alpha_{1}+\alpha_{2}+\alpha_{3})\beta)}\nonumber
\end{align}%
\begin{align}
\  &  \frac{\tilde C_{+}^{\text{(M)}}(\alpha_{1})C_{\text{M}}(\alpha_{1}%
-\beta^{-1}/2,\alpha_{2},\alpha_{3})}{\tilde C_{-}^{\text{(M)}}(\alpha
_{1})C_{\text{M}}(\alpha_{1}+\beta^{-1}/2,\alpha_{2},\alpha_{3})}=\label{F2}\\
&  -\dfrac{(1-\beta^{-2}+2\alpha_{1}\beta^{-1})^{2}\gamma(\beta^{-2}%
/2-(\alpha_{3}+\alpha_{1}-\alpha_{2})\beta^{-1})\gamma(\beta^{-2}%
/2-(\alpha_{1}+\alpha_{2}-\alpha_{3})\beta^{-1})}{\gamma^{2}(\beta
^{-2}-2\alpha_{1}\beta^{-1})\gamma(\beta^{-2}/2-(\alpha_{2}+\alpha_{3}%
-\alpha_{1})\beta^{-1})\gamma(2-3\beta^{-2}/2+(\alpha_{1}+\alpha_{2}%
+\alpha_{3})\beta^{-1})}\nonumber
\end{align}

First of all let's use these equations to recover explicit expressions for the
special structure constants $C_{\pm}^{\text{(M)}}(\alpha)$ and $\tilde C_{\pm
}^{\text{(M)}}(\alpha)$. Take $\alpha_{1}=\alpha_{3}=\alpha$ and $\alpha
_{2}=\beta/2$. With the use of (\ref{F1}) this results in
\begin{equation}
\left(  \ \frac{C_{+}^{\text{(M)}}(\alpha)}{C_{-}^{\text{(M)}}(\alpha
)}\right)  ^{2}=\dfrac{\gamma(2\alpha\beta)\gamma(2-\beta^{2}-2\alpha\beta
)}{\gamma(2-2\beta^{2}-2\alpha\beta)\gamma(\beta^{2}+2\alpha\beta)}\label{E1}%
\end{equation}
Similarly (\ref{F2}) gives
\begin{equation}
\left(  \ \frac{\tilde C_{+}^{\text{(M)}}(\alpha)}{\tilde C_{-}^{\text{(M)}%
}(\alpha)}\right)  ^{2}=\dfrac{\gamma(-2\alpha\beta^{-1})\gamma(2-\beta
^{-2}+2\alpha\beta^{-1})}{\gamma(2-2\beta^{-2}+2\alpha\beta^{-1})\gamma
(\beta^{2}-2\alpha\beta^{-1})}\label{E2}%
\end{equation}
Plugging (\ref{E1}) to (\ref{F1}) we arrive at a closed functional relation
for $C_{\text{M}}(\alpha_{1},\alpha_{2},\alpha_{3})$
\begin{align}
&  \ \frac{C_{\text{M}}(\alpha_{1}+\beta,\alpha_{2},\alpha_{3})}{C_{\text{M}%
}(\alpha_{1},\alpha_{2},\alpha_{3})}=\dfrac{\gamma(\beta^{2}+(\alpha
_{1}+\alpha_{2}-\alpha_{3})\beta)\gamma(\beta^{2}+(\alpha_{3}+\alpha
_{1}-\alpha_{2})\beta)}{\gamma((\alpha_{2}+\alpha_{3}-\alpha_{1})\beta
)\gamma(2-2\beta^{2}-(\alpha_{1}+\alpha_{2}+\alpha_{3})\beta)}\times
\label{CMshift1}\\
&  \ \ \ \ \left(  \gamma(\beta^{2}+2\alpha_{1}\beta)\gamma(2\beta^{2}%
+2\alpha_{1}\beta)\gamma(-1+2\beta^{2}+2\alpha_{1}\beta)\gamma(-1+3\beta
^{2}+2\alpha_{1}\beta)\right)  ^{-1/2}\nonumber
\end{align}
The second functional relation, which is combined from (\ref{F2}) and
(\ref{E2}) differs from the first one in the simple substitution
$\beta\rightarrow-\beta^{-1}$
\begin{align}
&  \frac{C_{\text{M}}(\alpha_{1}-\beta^{-1},\alpha_{2},\alpha_{3}%
)}{C_{\text{M}}(\alpha_{1},\alpha_{2},\alpha_{3})}=\dfrac{\gamma(\beta
^{-2}-(\alpha_{1}+\alpha_{2}-\alpha_{3})\beta^{-1})\gamma(\beta^{-2}%
-(\alpha_{3}+\alpha_{1}-\alpha_{2})\beta^{-1})}{\gamma(-(\alpha_{2}+\alpha
_{3}-\alpha_{1})\beta^{-1})\gamma(2-2\beta^{-2}+(\alpha_{1}+\alpha_{2}%
+\alpha_{3})\beta^{-1})}\times\label{CMshift2}\\
&  \ \ \ \ \ \left(  \gamma(\beta^{-2}-2\alpha_{1}\beta^{-1})\gamma
(2\beta^{-2}-2\alpha_{1}\beta^{-1})\gamma(2\beta^{-2}-2\alpha_{1}\beta
^{-1}-1)\gamma(3\beta^{-2}-2\alpha_{1}\beta^{-1}-1)\right)  ^{-1/2}\nonumber
\end{align}

Since all the bootstrap calculations above are apparently analytic in $\beta$
and $\alpha$, it's not a big surprise, that these functional relations (up to
the factors which depend on the parameter $\alpha_{1}$ only and not on the
combinations of different $\alpha$, and therefore are sensitive to the
normalization of the operators) are precisely the analytic continuation of the
similar functional relations (\ref{CLshift}) in the Liouville field theory
under
\begin{align}
b  &  \rightarrow-i\beta\nonumber\\
b^{-1}  &  \rightarrow i\beta^{-1}\label{cont}\\
a  &  \rightarrow i\alpha\nonumber
\end{align}
However, for the reasons to be explained below, an attempt to construct the
real $\beta$ solution through such continuation fails and the solution
presented in the next section is not an analytic continuation of the Liouville
three-point function (\ref{CL}).

\section{GMM three-point function}

For real and incommensurable $\beta$ and $\beta^{-1}$ the solution to the
system (\ref{CMshift1}, \ref{CMshift2}) is unique. It can be expressed in
terms of the same special function $\Upsilon(x)=\Upsilon_{\beta}(x)$ as in
eq.(\ref{CL}) but now with the parameter $\beta$ instead of $b$
\begin{align}
&  C_{\text{M}}(\alpha_{1},\alpha_{2},\alpha_{3})=\label{CM}\\
&  \frac{A\Upsilon(\alpha_{1}+\alpha_{2}-\alpha_{3}+\beta)\Upsilon(\alpha
_{2}+\alpha_{3}-\alpha_{1}+\beta)\Upsilon(\alpha_{3}+\alpha_{1}-\alpha
_{2}+\beta)\Upsilon(2\beta-\beta^{-1}+\alpha_{1}+\alpha_{2}+\alpha_{3}%
)}{\left[  \Upsilon(2\alpha_{1}+\beta)\Upsilon(2\alpha_{1}+2\beta-\beta
^{-1})\Upsilon(2\alpha_{2}+\beta)\Upsilon(2\alpha_{2}+2\beta-\beta
^{-1})\Upsilon(2\alpha_{3}+\beta)\Upsilon(2\alpha_{3}+2\beta-\beta
^{-1})\right]  ^{1/2}}\nonumber
\end{align}
where the normalization factor $A$
\begin{equation}
A=\frac{\beta^{\beta^{-2}-\beta^{2}-1}\left[  \gamma(\beta^{2})\gamma
(\beta^{-2}-1)\right]  ^{1/2}}{\Upsilon(\beta)}\label{A}%
\end{equation}
is determined from the normalization requirement $C_{\text{M}}(0,\alpha
,\alpha)=1$ (we insist to interpret $\Phi_{0}$ as the identity operator in GMM).

In particular, the truncated operator product expansions (\ref{Phi12Phia})
read explicitly
\begin{align}
&  \Phi_{1,2}\Phi_{\alpha}=\left[  \frac{\gamma(\beta^{2})\gamma(2\alpha
\beta+2\beta^{2}-1)}{\gamma(2\beta^{2}-1)\gamma(\beta^{2}+2\alpha\beta
)}\right]  ^{1/2}\left[  \Phi_{\alpha+\beta/2}\right]  +\left[  \frac
{\gamma(\beta^{2})\gamma(2\alpha\beta+\beta^{2}-1)}{\gamma(2\beta^{2}%
-1)\gamma(2\alpha\beta)}\right]  ^{1/2}\left[  \Phi_{\alpha-\beta/2}\right]
\nonumber\\
&  \Phi_{2,1}\Phi_{\alpha}=\label{Phi12x}\\
&  \left[  \frac{\gamma(\beta^{-2})\gamma(2\beta^{-2}-2\alpha\beta^{-1}%
-1)}{\gamma(2\beta^{-2}-1)\gamma(\beta^{-2}-2\alpha\beta^{-1})}\right]
^{1/2}\left[  \Phi_{\alpha-\beta^{-1}/2}\right]  +\left[  \frac{\gamma
(\beta^{-2})\gamma(\beta^{-2}-2\alpha\beta^{-1}-1)}{\gamma(2\beta
^{-2}-1)\gamma(-2\alpha\beta^{-1})}\right]  ^{1/2}\left[  \Phi_{\alpha
+\beta/2}\right] \nonumber
\end{align}

I have verified for many particular examples that once specified for the
allowed degenerate values of $\alpha_{i}=\alpha_{m_{i},n_{i}}$ expression
(\ref{CM}) reduces precisely to the Dotsenko-Fateev products of gamma
functions. It is important to this order that the set of the three
$\alpha_{m_{i},n_{i}}$ in the structure constant satisfy the fusion rules
(\ref{ff}). Otherwise the expression doesn't always take care of the vanishing
of the structure constant. Often it produces certain finite numbers, whose
meaning and possible interpretation remains unclear to me. It would be
important to prove the identity of (\ref{CM}) and the Dotsenko-Fateev
expressions for the allowed degenerate values of $\alpha_{i}$.

\begin{figure}
[ptb]
\begin{center}
\includegraphics[
height=3.1488in,
width=6.1488in
]%
{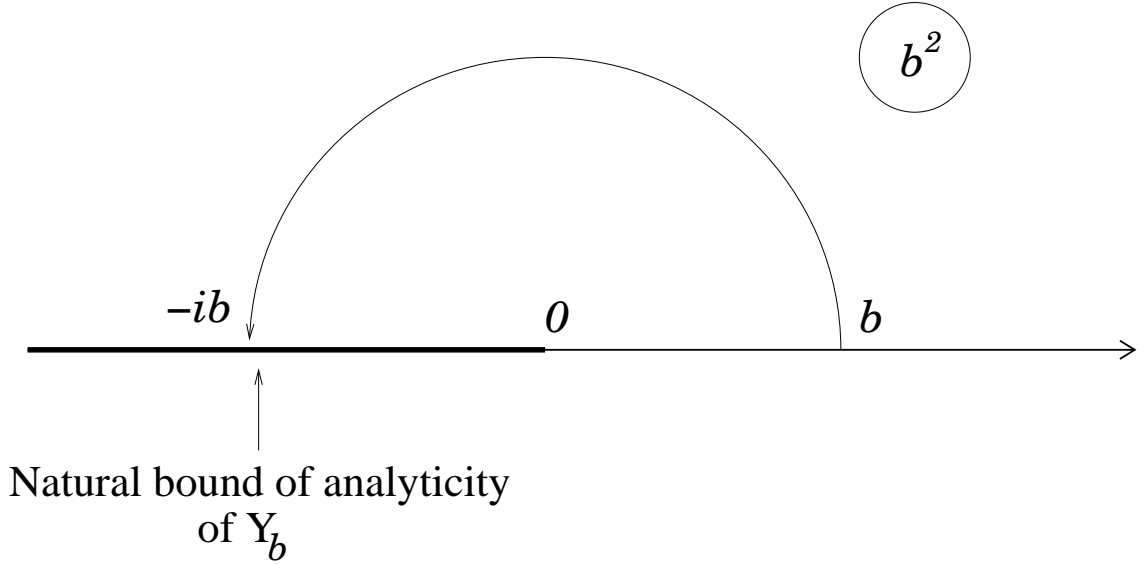}%
\caption{Analytic properties of $\Upsilon_{b}(x)$ as a function of $b^{2}$}%
\label{fig1}%
\end{center}
\end{figure}

The solution (\ref{CM}) is not an analytic continuation of the Liouville
three-point function (\ref{CL}). The point is that the function $\Upsilon
_{b}(x)$ as a function of its parameter $b$ is analytic in the whole complex
plane of $b^{2}$ except for the negative part of the real axis, where it meets
with a natural bound of analyticity (see fig.\ref{fig1}). In the next section
we comment a little more about the analytic relation between the solutions
(\ref{CL}) and (\ref{CM}).%

\section{Analytic continuation in $b^{2}$}

\begin{figure}
[ptb]
\begin{center}
\includegraphics[
height=3.6893in,
width=6.1488in
]%
{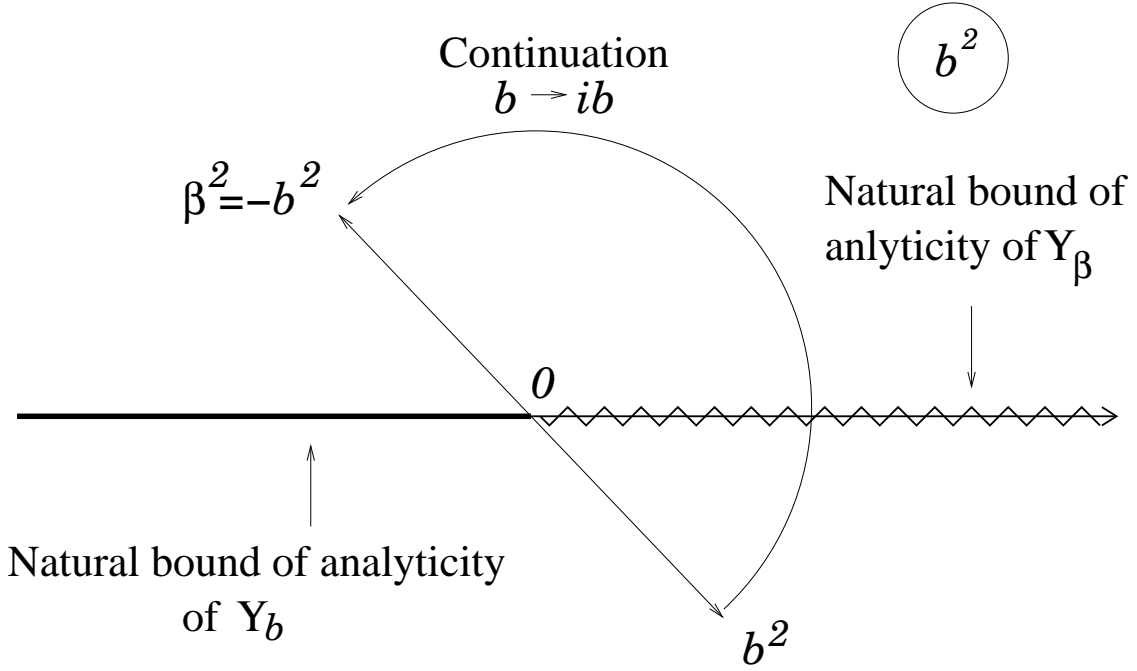}%
\caption{Analytic continuation of $\Upsilon_{b}$ (with $\operatorname*{Im}%
b^{2}<0$) to $\Upsilon_{\beta}$ with $\beta=ib $}%
\label{fig2}%
\end{center}
\end{figure}

As it was mentioned in the previous section, $C_{\text{L}}(a_{1},a_{2}%
,a_{3}|b)$ cannot be analytically continued to pure imaginary values of the
parameter $b $. For the same reason $C_{\text{M}}(\alpha_{1},\alpha_{2}%
,\alpha_{3}|\beta)$ doesn't allow an analytic continuation to purely imaginary
values of $\beta$. At the same time $C_{\text{L}}(a_{1},a_{2},a_{3}|b)$ and
$C_{\text{M}}(-ia_{1},-ia_{2},-ia_{3}|ib) $ satisfy the same set of functional
shift relations (up to some normalization dependent multipliers), which allow
to reconstruct uniquely $C_{\text{M}}(-ia_{1},-ia_{2},-ia_{3}|ib)$ or
$C_{\text{L}}(a_{1},a_{2},a_{3}|b) $ at any positive irrational or negative
irrational value of $b^{2}$ respectively. To clarify the analytic relation
between these two functions, let's compare them at complex values of $b^{2}$
where both are well defined. At complex values of $b$ the shift relations
determine the solution only up to an elliptic function with periods $b$ and
$b^{-1}$. Take $C_{\text{L}}(a_{1},a_{2},a_{3}|b)$ at certain complex value of
$b^{2}$ in the lower half-plane, as it is shown in fig.\ref{fig2}, so that it
can be safely continued to the point $\beta=ib$ and compared there with
$C_{\text{M}}(-ia_{1},-ia_{2},-ia_{3}|ib) $.%

Consider the ratio
\begin{equation}
R(a_{1},a_{2},a_{3}|b)=\frac{C_{\text{M}}(-ia_{1},-ia_{2},-ia_{3}%
|ib)}{C_{\text{L}}(a_{1},a_{2},a_{3}|b)}\label{R}%
\end{equation}
In this ratio the $\Upsilon$-functions entering the GMM and Liouville
three-point functions combine together to theta-functions due to the following
identity (see Appendix B for the proof)
\begin{equation}
\Upsilon_{b}(x)\Upsilon_{ib}(-ix+ib)=e^{i\pi(b-b^{-1}-2x)^{2}/8}\frac
{\theta_{1}(\pi xb^{-1}|h)}{h^{1/4}\theta_{3}(0|h)}\label{complementarity}%
\end{equation}
where $\theta_{1}(x|h)$ and $\theta_{3}(x|h)$ are usual elliptic
theta-functions and $h=\exp(i\pi b^{-2}).$ It is implied in
eq.(\ref{complementarity}) that $b^{2}$ has negative imaginary part, so that
$\Upsilon_{b}$ can be continued safely to $\Upsilon_{ib}$ avoiding the
negative part of real axis of $b^{2}$, as it is shown in fig.\ref{fig2}. The
result is
\begin{equation}
\ R(a_{1},a_{2},a_{3})=\frac{\left(  \pi\mu\gamma(b^{2})\right)
^{(a_{1}+a_{2}+a_{3}-Q)/b}\left[  \gamma(-b^{2})\gamma(-b^{-2}-1)\right]
^{1/2}}{b^{4}\prod_{i=1}^{3}\left[  \gamma(2a_{i}b-b^{2}-1)\gamma(2a_{i}%
b^{-1}-b^{-2})\right]  ^{1/2}}T(a_{1},a_{2},a_{3})\label{analytic}%
\end{equation}
where
\begin{align}
&  T(a_{1},a_{2},a_{3})=\label{Telliptic}\\
&  \ \frac{\theta_{1}\left(  \pi b^{-1}(a_{1}+a_{2}-a_{3})\right)  \theta
_{1}\left(  \pi b^{-1}(a_{2}+a_{3}-a_{1})\right)  \theta_{1}\left(  \pi
b^{-1}(a_{3}+a_{1}-a_{2})\right)  \theta_{1}\left(  \pi b^{-1}(a_{1}%
+a_{2}+a_{3})\right)  }{\pi b^{-1}\theta_{1}^{\prime}\left(  0\right)
\theta_{1}\left(  2\pi b^{-1}a_{1}\right)  \theta_{1}\left(  2\pi b^{-1}%
a_{2}\right)  \theta_{1}\left(  2\pi b^{-1}a_{3}\right)  }\nonumber
\end{align}
(all $\theta$-functions has the same modulus $h$) is verified to be an
elliptic function in all three $a_{i}$ with periods $b$ and $b^{-1}$. Thus,
for complex $b^{2}$ the two solutions $C_{\text{L}}(a_{1},a_{2},a_{3}|b)$ and
$C_{\text{M}}(-ia_{1},-ia_{2},-ia_{3}|ib)$ of the same system of functional
relations differ, as it should be, in a double periodic function (up to the
``leg-factor'' multiplier in (\ref{analytic}) which is due to different
normalization of the primary fields in GMM and Liouville field theory).

\section{Minimal gravity}

Let's now turn to the minimal gravity, induced by the GMM $\mathcal{M}%
_{\beta^{2}}$ with the total matter central charge (\ref{cM}). As it is widely
known, the central charge balance (\ref{cbalance}) requires the Liouville
parameter $b$ be equal the GMM parameter $\beta$ (of course there is another
solution $b=\beta^{-1}$, we choose the first one in order to stay with $b<1$).
Therefore from now on we'll use single notation $b$ for this parameter in both
Liouville and GMM, substituting
\begin{equation}
\beta=b\label{betab}%
\end{equation}
in all GMM-related expressions. Moreover, the ``dressing parameter'' $a$ in
eq.(\ref{composite}) in MG is simply related to the parameter $\alpha$ of the
matter primary field
\begin{equation}
a=\alpha+b\label{aalpha}%
\end{equation}
Again, there is a second possibility $a=b^{-1}-\alpha$, but this second choice
doesn't offer anything new, since the analytic expressions for LFT correlation
functions allow to identify these two possible dressings up to the
normalization \cite{ZZ1}
\begin{equation}
\exp(2a\phi)=D_{\text{L}}(a)\exp(2(Q-a)\phi)\label{reflection}%
\end{equation}
with $D_{\text{L}}(a)$ from eq.(\ref{Da}). We'll denote the corresponding
(formal for generic $\alpha$) dressed operator
\begin{equation}
U_{a}=\Phi_{a-b}e^{2a\phi}\label{MGdressing}%
\end{equation}
and the ``reflected'' one
\begin{equation}
\tilde U_{a}=U_{Q-a}=\Phi_{b^{-1}-a}e^{2(Q-a)\phi}\label{Utwiddle}%
\end{equation}
Notice, that due to the identification (\ref{q-a})
\begin{equation}
\tilde U_{a}=D_{\text{L}}(Q-a)U_{a}\label{Ureflected}%
\end{equation}

In the minimal gravity the three-point function reads simply as
\begin{equation}
C^{\text{(MG)}}(a_{1},a_{2},a_{3})=\left\langle C\bar C(x_{1})\ldots
CC(x_{3})\right\rangle _{\text{gh}}\left\langle U_{a_{1}}(x_{1})U_{a_{2}%
}(x_{2})U_{a_{3}}(x_{3})\right\rangle _{\text{MG}}\label{CMG}%
\end{equation}
The coordinate dependence cancels out between the two multipliers in the right
hand side and we arrive, as expected, at the three-point MG correlation
number
\begin{equation}
C^{\text{(MG)}}(a_{1},a_{2},a_{3})=C_{\text{M}}(a_{1}-b,a_{2}-b,a_{3}%
-b|b)C_{\text{L}}(a_{1},a_{2},a_{3}|b)\label{CMG2}%
\end{equation}
In this product all $\Upsilon$-functions dependent on the combinations of the
parameters $a_{i}$ cancel out and we're left with (up to an overall
multiplier) a product of factors dependent on each individual parameter
$a_{i}$ (sometimes called the ``leg-factors'')
\begin{equation}
C^{\text{(MG)}}(a_{1},a_{2},a_{3})=\left(  \pi\mu\gamma(b^{2})\right)
^{(Q-\sum_{i=1}^{3}a_{i})/b}\left[  \frac{\gamma(b^{2})\gamma(b^{-2}-1)}%
{b^{2}}\right]  ^{1/2}\prod_{i=1}^{3}F_{\text{MG}}(a_{i})\label{CMGproduct}%
\end{equation}
where we have chosen the following normalization of the leg-factors
\begin{equation}
F_{\text{MG}}(a)=\left[  \gamma(2ab-b^{2})\gamma(2ab^{-1}-b^{-2})\right]
^{1/2}\label{leg}%
\end{equation}

This factorized form of the MG three point function has been established
already (and long ago) in the matrix model approach \cite{DiFrancesco} as well
as within a somewhat different approach to Liouville field theory
\cite{GoulianLi}.

Of course, technically it is much easier not to solve separately the
functional relations (\ref{CLshift}) and (\ref{CMshift1}, \ref{CMshift2}) for
each Liouville and GMM correlation functions but rather combine them as the
functional relations for the product. This way one avoids completely any need
in complicated special functions $\Upsilon$. Combining (\ref{CLshift}) and
(\ref{CMshift1}, \ref{CMshift2}) with the relation (\ref{aalpha}) taken into
account one finds the equations
\begin{align}
\frac{C_{\text{MG}}(a_{1}+b,a_{2},a_{3})}{C_{\text{MG}}(a_{1},a_{2},a_{3})}
&  =\frac{(2a_{1}b^{-1}-b^{-2})(2a_{1}b^{-1}+1-b^{-2})}{\pi\mu\gamma(b^{2}%
)}\ \left(  \frac{\gamma(2a_{1}b+b^{2})}{\gamma(2a_{1}b-b^{2})}\right)
^{1/2}\label{CMGshift}\\
\frac{C_{\text{MG}}(a_{1}+b^{-1},a_{2},a_{3})}{C_{\text{MG}}(a_{1},a_{2}%
,a_{3})}  &  =\frac{(2a_{1}b-b^{2}+1)(2a_{1}b-b^{2})}{\pi\tilde\mu
\gamma(b^{-2})}\left(  \frac{\gamma(2a_{1}b^{-1}+b^{-2})}{\gamma(2a_{1}%
b^{-1}-b^{-2})}\right)  ^{1/2}\nonumber
\end{align}
which are enough to restore the factorized form of (\ref{CMGproduct}) up to
the $\alpha$-independent overall normalization. We have chosen in this study
the more complicated scheme of separate calculation of the GMM structure
function to reveal more information about individual contributions of the
Liouville and GMM degrees of freedom to the minimal gravity result
(\ref{CMGproduct}).

Finally, we have to specialize the analytic expression for the degenerate
values of the matter field parameters. At this step the multiplier
$f_{m_{1},m_{2},m_{2}}f_{n_{1},n_{2},n_{2}}$ should be added to ensure the
fusion rules for the GMM degenerate fields. We arrive at the following
expression for the minimal gravity three-point correlation numbers
\begin{align}
C_{(m_{1},n_{1})(m_{2},n_{2})(m_{3},n_{3})}^{\text{(MG)}}  &  =f_{m_{1}%
,m_{2},m_{2}}f_{n_{1},n_{2},n_{2}}\times\label{CMGmmm}\\
&  \left(  \pi\mu\gamma(b^{2})\right)  ^{(Q-\sum_{i=1}^{3}a_{m_{i},n_{i}})/b}
\left[  \frac{\gamma(b^{2})\gamma(b^{-2}-1)}{b^{2}}\right]  ^{1/2}\prod
_{i=1}^{3}F_{\text{MG}}(a_{m_{i},n_{i}})\nonumber
\end{align}

As it was mentioned above, the Liouville three-point function is the
non-normalized one. To arrive at the normalized expression we have still to
divide it by the Liouville partition function. This will be done in the next section.

\section{Two-point function and normalization}

Contrary to the three-point correlation numbers, it is not that easy to carry
out directly the two-point ones. In the direct calculation the problem of
fixing the remaining gauge symmetry on the sphere should be treated
separately. To avoid this complicated task we prefer here to use the identity
\begin{equation}
-\frac\partial{\partial\mu}\left\langle U_{a}U_{a}\right\rangle _{\text{g}%
}=\left\langle U_{a}U_{a}U_{b}\right\rangle _{\text{g}}\label{dmu}%
\end{equation}
which follows directly from the action (\ref{Ag}) and the Liouville Lagrangian
(\ref{AL}). This results in the following expression for the (non-normalized)
two-point correlation number in MG
\begin{equation}
\left\langle U_{a}U_{a}\right\rangle _{\text{MG}}=\left(  \pi\mu\gamma
(b^{2})\right)  ^{(Q-2a)/b}\frac{F_{\text{MG}}^{2}(a)}{\pi(2a-Q)}%
\label{twopoint}%
\end{equation}
Similar relation
\begin{equation}
-\frac{\partial^{3}}{\partial\mu^{3}}Z_{\text{L}}=C_{\text{MG}}(b,b,b)=\left(
\pi\mu\gamma(b^{2})\right)  ^{Q/b}\frac{\gamma(2-b^{-2})}{\pi^{3}\gamma
(b^{2})\mu^{3}b}\label{d3mu}%
\end{equation}
can be used to restore the Liouville partition function \cite{DO}
\begin{equation}
Z_{\text{L}}=\left(  \pi\mu\gamma(b^{2})\right)  ^{Q/b}\frac{(1-b^{2})}%
{\pi^{3}\gamma(b^{2})\gamma(b^{-2})Q}\label{ZL}%
\end{equation}

With this expression the normalized two- and three-point correlation numbers
in GMM read
\begin{equation}
\frac{\left\langle U_{a}U_{a}\right\rangle _{\text{MG}}}{Z_{\text{L}}}%
=\frac{\left(  \pi\mu\gamma(b^{2})\right)  ^{-2a/b}\pi^{2}\gamma(b^{2}%
)\gamma(b^{-2})Q}{(1-b^{2})(2a-Q)}F_{\text{MG}}^{2}(a)\label{twopointn}%
\end{equation}
and
\begin{equation}
\frac{C^{\text{(MG)}}(a_{1},a_{2},a_{3})}{Z_{\text{L}}}=\left(  \pi\mu
\gamma(b^{2})\right)  ^{-\sum_{i=1}^{3}a_{i}/b}\frac{\pi^{3}(1+b^{2})\left[
-\gamma^{3}(b^{2})\gamma^{3}(b^{-2})\right]  ^{1/2}}{(1-b^{2})^{2}}\prod
_{i=1}^{3}F_{\text{MG}}(a_{i})\label{threepointn}%
\end{equation}

Expressions (\ref{twopointn}) and (\ref{threepointn}), together with the
representation (\ref{CM}) for the structure constants in GMM, are the main
results of the presented study.

\section{Matrix models}

Let me quote here two results coming from the matrix model approach.

\textbf{1. Scaling Lee-Yang model }(SLYM) originates from the minimal model
$\mathcal{M}_{2,5}$ of central charge $-22/5$. It contains only one
non-trivial primary field $\Phi=\Phi_{1,2}=\Phi_{1,3}$ of dimension $-1/5$.
This relevant field gives rise to the only possible perturbation (with pure
imaginary coupling constant), a massive field theory called SLYM. It seems
natural to call the perturbed MG based on this CFT the gravitational Lee-Yang
model (GLY). It can be conventionally encoded into the formal action
\begin{equation}
A_{\text{GLY}}(\lambda,\mu)=A_{\text{MG}}(2/5)+\mu\int e^{2\sqrt{2/5}\phi
}+\frac{ih}{2\pi}\int\Phi_{1,2}e^{3\sqrt{2/5}\phi}\label{GLY}%
\end{equation}
The two- and three-point correlation functions of the dressed perturbing field
$U_{1,2}=\Phi_{1,2}e^{3\sqrt{2/5}\phi}$ come out from (\ref{twopointn}) and
(\ref{threepointn}) as
\begin{align}
\frac{\left\langle U_{1,2}U_{1,2}\right\rangle _{\text{GLY}}}{Z_{\text{L}}}
&  =\frac{105(2\pi L_{\text{GLY}})^{2}}4\label{GLY23}\\
\frac{\left\langle U_{1,2}U_{1,2}U_{1,2}\right\rangle _{\text{GLY}}%
}{Z_{\text{L}}}  &  =\frac{105(2\pi L_{\text{GLY}})^{3}}8\nonumber
\end{align}
where
\begin{equation}
L_{\text{GLY}}=\frac{-il}{(\pi\mu)^{3/2}}\label{LGLY}%
\end{equation}
and
\begin{equation}
l=\frac{\gamma^{1/2}(4/5)}{4\gamma(2/5)}\label{l}%
\end{equation}

One of the most early results of the matrix model calculations states that an
appropriate scaling limit of the exactly solvable one-matrix model can be
interpreted as SLYM coupled to quantum gravity, i.e., GLY \cite{Staudacher}.
In particular, from this identification the sphere partition sum
$Z_{\text{GLY}}(t,x)$ (parameters $x$ and $t$ come from matrix model and
identified, up to normalization, with the cosmological constant $\mu$ and the
$\phi$-coupling $\lambda$ respectively) comes out in the following explicit
form. Let
\begin{equation}
\frac{\partial^{2}}{\partial t^{2}}Z_{\text{GLY}}(t,x)=u(t,x)\label{dZGLY}%
\end{equation}
Function $u(t,x)$ is an appropriate solution to the simple algebraic equation
\begin{equation}
t=u^{3}-xu\label{uGLY}%
\end{equation}
Equations (\ref{ZGLY}) and (\ref{uGLY}) give the following expansion
\begin{align}
Z_{\text{GLY}}(t,x) &  =x^{7/2}\sum_{\substack{n=0 \\n\neq1 }}^{\infty}%
\frac{\Gamma(3n/2-7/2)\left(  tx^{-3/2}\right)  ^{n}}{2n!\Gamma(n/2-1/2)}%
\label{ZGLY}\\
\  &  =Z_{\text{GLY}}(0,x)\left(  1+\frac{105}8\left(  tx^{-3/2}\right)
^{2}-\frac{35}{16}\left(  tx^{-3/2}\right)  ^{3}-\frac{105}{128}\left(
tx^{-3/2}\right)  ^{4}+\ldots\right) \nonumber
\end{align}
The second and third order coefficients in this expansion are consistent with
the two- and three-point functions (\ref{GLY23}) provided
\begin{equation}
tx^{-3/2}=lh\label{lh}%
\end{equation}
with $l$ from (\ref{l}). Moreover, the four-point function is expected to be
\begin{equation}
\frac{\left\langle U_{1,2}U_{1,2}U_{1,2}U_{1,2}\right\rangle _{\text{GLY}}%
}{Z_{\text{L}}}=-\frac{315}{16}L_{\text{GLY}}^{4}\label{GLY4}%
\end{equation}

\textbf{2. Generalized minimal model }$\mathcal{M}_{b^{2}}$ coupled to gravity
and perturbed by $\Phi_{1,3}$. Here we consider the region $b^{2}>1/3$. In
this region it turns to be consistent with matrix models to choose the
``reflected'' (as compared to (\ref{MGdressing})) dressing $\tilde
U_{1,3}=\Phi_{1,3}e^{2(b^{-1}-b)\phi}$, the action having the form
\begin{equation}
A(\lambda,\mu)=A_{\text{MG}}(b^{2})+\mu\int e^{2b\phi}+\frac\lambda{2\pi}%
\int\Phi_{1,3}e^{2(b^{-1}-b)\phi}\label{pMG}%
\end{equation}
Putting $a=b^{-1}-b$ in the expressions (\ref{twopointn}) and
(\ref{threepointn})
\begin{align}
\frac{\left\langle \tilde U_{1,3}\tilde U_{1,3}\right\rangle _{\text{MG}}%
}{Z_{\text{L}}} &  =-\frac{(g-1)g(g+1)}{(g-3)}(2\pi L_{g})^{2}\label{MG23}\\
\frac{\left\langle \tilde U_{1,3}\tilde U_{1,3}\tilde U_{1,3}\right\rangle
_{\text{MG}}}{Z_{\text{L}}} &  =(g-1)g(g+1)(2\pi L_{g})^{3}\nonumber
\end{align}
where the following notation is introduced
\begin{equation}
L_{g}=-\frac{\gamma(g-1)[-\gamma(b^{2})\gamma(2-3b^{2})]^{1/2}}{2(g-2)\left(
\pi\mu\gamma(b^{2})\right)  ^{g-1}}\label{Lg}%
\end{equation}

The $O(n)$ model (a gas of self-avoiding loops) on a random lattice
\cite{KostovOn} is interpreted in the continuum limit of GMM with the central
charge
\begin{equation}
c_{g}=13-6\left(  g+g^{-1}\right) \label{cg}%
\end{equation}
related to $n$ (the statistical weight of each loop) as
\begin{equation}
n=-2\cos(\pi g)\label{ng}%
\end{equation}
This allows to identify
\begin{equation}
g=b^{-2}\label{gb}%
\end{equation}
In what follows it is convenient also to use the parameter
\begin{equation}
p=\frac{b^{2}}{1-b^{2}}=\frac1{g-1}\label{p}%
\end{equation}

A recent (unpublished) result by I.Kostov gives explicitly the sphere
partition function of gravitational ``massive'' (or dilute) loop gas. Let
again $x$ be the ``cosmological constant'', i.e. the lattice version of $\mu$,
and $t$ be related to the ``mass'' of the loop, i.e., the deviation of the
loop length activity from its critical value (the value where loops ``blow
up''). Let also $Z(t,x)$ be the partition function of the perturbed MG
(\ref{pMG}). Then
\begin{equation}
u=-(g-1)Z_{xx}\label{ugZ}%
\end{equation}
solves the following transcendental equation \cite{Kostov}
\begin{equation}
u^{p}+tu^{p-1}=x\label{ieq}%
\end{equation}
Thus the partition sum is expanded as the following perturbative series
\begin{align}
Z &  =tx^{2}+x^{g+1}\sum_{\substack{n=0 \\n\neq1 }}^{\infty}\frac
{\Gamma(g(n-1)-n-1)}{n!\Gamma(g(n-1)-2n+2)}\left(  tx^{1-g}\right)
^{n}\label{Zg13}\\
\  &  =x^{g+1}\left(  \frac{-1}{(g-1)g(g+1)}+tx^{1-g}+\frac{\left(
tx^{1-g}\right)  ^{2}}{2(g-3)}+\frac{\left(  tx^{1-g}\right)  ^{3}}%
6+\frac{(g-2)\left(  tx^{1-g}\right)  ^{4}}8+\ldots\right) \nonumber
\end{align}
Again, this is consistent with the expressions (\ref{MG23}) provided
\begin{equation}
tx^{1-g}=2\pi L_{g}\lambda\label{tlambda}%
\end{equation}
and also allows to predict the four-point function
\begin{equation}
\frac{\left\langle \tilde U_{1,3}\tilde U_{1,3}\tilde U_{1,3}\tilde
U_{1,3}\right\rangle _{\text{MG}}}{Z_{\text{L}}}=-3(g-2)(g-1)g(g+1)(2\pi
L_{g})^{4}\label{MG4}%
\end{equation}

Notice also that (\ref{Zg13}) reduces to (\ref{ZGLY}) at $g=5/2$ (i.e., for
$b^{2}=2/5$), although the defining equations (\ref{uGLY}) and (\ref{ieq}) are
apparently different. This is as it should be, since GLY is the particular
case of the $\Phi_{1,3}$-perturbed MG for $b^{2}=2/5$. In particular
\begin{equation}
L_{\text{GLY}}=L_{5/2}\label{L52}%
\end{equation}

\textbf{Acknowledgments}

This work is for Galina Gritsenko. I'm also grateful to A.Belavin and
Alexander Zamolodchikov for encouraging interest to the work. Special
gratitude to Ivan Kostov for useful discussions and sharing the results of
\cite{Kostov} prior to publication. A particular acknowledgment to I.Kostov
and V.Petkova for their kind permission to enjoy the early acquaintance with
their forthcoming paper \cite{KP}. The work was supported by the European
Committee under contract HPRN-CT-2002-00325.

\appendix

\section{Functional relations}

In this appendix we repeat the standard bootstrap derivation of the functional
relations (\ref{F1},\ref{F2}) (and in fact, similar Liouville relations
(\ref{CLshift}), see \cite{Teschner1}). This derivation is based on the
truncated operator product expansions (\ref{Phi12Phia}) which, in order, are
the consequences of the vanishing of the corresponding level two null-vectors
(\ref{diff12}).

Here we consider only the first relation (\ref{F1}), having in mind that all
the considerations remain valid if we replace $\Phi_{1,2}\rightarrow\Phi
_{2,1}$ and $\beta\rightarrow-\beta^{-1}$. Consider the following four-point
function
\begin{equation}
G=G\left(
\begin{array}
[c]{cc}%
\Delta_{1} & \Delta_{3}\\
\Delta_{2} & \Delta_{1,2}%
\end{array}
|
\begin{array}
[c]{cc}%
x_{1} & x_{3}\\
x_{2} & x_{4}%
\end{array}
\right)  =\left\langle \Phi_{\alpha_{1}}(x_{1})\Phi_{\alpha_{2}}(x_{2}%
)\Phi_{\alpha_{3}}(x_{3})\Phi_{1,2}(x_{4})\right\rangle _{\text{GMM}%
}\label{fourp}%
\end{equation}
As a consequence of conformal invariance this function has the form
\cite{BPZ}
\begin{equation}
G=\frac{g(x,\bar x)}{(x_{34}\bar x_{34})^{2\Delta_{1,2}}(x_{13}\bar
x_{13})^{\Delta_{1}+\Delta_{3}-\Delta_{1,2}-\Delta_{2}}(x_{23}\bar
x_{23})^{\Delta_{2}+\Delta_{3}-\Delta_{1,2}-\Delta_{1}}(x_{12}\bar
x_{12})^{\Delta_{1,2}+\Delta_{1}+\Delta_{2}-\Delta_{3}}}\label{fourpreduced}%
\end{equation}
where $\Delta_{1,2}=-1/2+3\beta^{2}/4$ and for the sake of brevity we denoted
$\Delta_{i}=\Delta_{\alpha_{i}}=\alpha_{i}(\alpha_{i}-q)$ for $i=1,2,3$.
Function $g$ depends only on the projective invariant
\begin{equation}
x=\frac{x_{41}x_{23}}{x_{43}x_{21}}\label{projinv}%
\end{equation}
As a consequence of the vanishing (\ref{diff12}) of the invariant vector the
following differential equation holds \cite{BPZ}
\begin{equation}
\ \beta^{-2}\frac{\partial^{2}}{\partial x_{4}^{2}}G=\sum_{i=1}^{3}\left(
\frac{\Delta_{i}}{(x_{4}-x_{i})^{2}}+\frac1{x_{4}-x_{i}}\frac\partial{\partial
x_{i}}\right)  G\label{partial}%
\end{equation}
With (\ref{fourpreduced}) this equations is reduced to an ordinary
differential one for $g(x,\bar x)$
\begin{equation}
-\beta^{-2}g_{xx}+\frac{2x-1}{x(1-x)}g_{x}+\left(  \frac{\Delta_{1}}{x^{2}%
}+\frac{\Delta_{2}}{(1-x)^{2}}+\frac{(\Delta_{1,2}+\Delta_{1}+\Delta
_{2}-\Delta_{3})}{x(1-x)}\right)  g=0\label{gdiff}%
\end{equation}
Of course, similar ``left'' equation with $x\rightarrow\bar x$ also holds.

Two independent solutions to this equation can be chosen either as the
``$s$-channel'' four-point blocks \cite{BPZ}
\begin{align}
\mathcal{F}\left(
\begin{array}
[c]{cc}%
\beta/2 & \alpha_{2}\\
\alpha_{1} & \alpha_{3}%
\end{array}
|\alpha_{1}+\beta/2|x\right)   &  =\mathcal{F}_{+}^{(s)}(x)=x^{\beta\alpha
_{1}}(1-x)^{\beta\alpha_{2}}{}_{2}F{}_{1}(A,B,C,x)\label{sblocks}\\
\mathcal{F}\left(
\begin{array}
[c]{cc}%
\beta/2 & \alpha_{2}\\
\alpha_{1} & \alpha_{3}%
\end{array}
|\alpha_{1}-\beta/2|x\right)   &  =\mathcal{F}_{-}^{(s)}(x)=x^{1-\beta
\alpha_{1}-\beta^{2}}(1-x)^{1-\beta\alpha_{2}-\beta^{2}}{}_{2}F{}%
_{1}(1-A,1-B,2-C,x)\nonumber
\end{align}
(which have a diagonal monodromy near $x=0$) or as the ``$u$-channel'' ones
\begin{align}
\mathcal{F}\left(
\begin{array}
[c]{cc}%
\beta/2 & \alpha_{1}\\
\alpha_{2} & \alpha_{3}%
\end{array}
|\alpha_{2}+\beta/2|x\right)   &  =\mathcal{F}_{+}^{(u)}(x)=x^{\beta\alpha
_{1}}(1-x)^{\beta\alpha_{2}}{}_{2}F{}_{1}(A,B,1+A+B-C,1-x)\label{ublocks}\\
\mathcal{F}\left(
\begin{array}
[c]{cc}%
\beta/2 & \alpha_{1}\\
\alpha_{2} & \alpha_{3}%
\end{array}
|\alpha_{2}-\beta/2|x\right)   &  =\mathcal{F}_{-}^{(u)}(x)\nonumber\\
&  =x^{1-\beta\alpha_{1}-\beta^{2}}(1-x)^{1-\beta\alpha_{2}-\beta^{2}}{}%
_{2}F{}_{1}(1-A,1-B,1+C-A-B,1-x)\nonumber
\end{align}
with a diagonal monodromy near $x=1$. Here
\begin{align}
A  &  =-1+3\beta^{2}/2+(\alpha_{1}+\alpha_{2}+\alpha_{3})\beta\nonumber\\
B  &  =\beta^{2}/2+(\alpha_{1}+\alpha_{2}-\alpha_{3})\beta\label{ABC}\\
C  &  =\beta^{2}+2\alpha_{1}\beta\nonumber
\end{align}
Standard transformations relate these two pairs of solutions
\begin{equation}
\left(
\begin{array}
[c]{c}%
\mathcal{F}_{+}^{(s)}(x)\\
\mathcal{F}_{-}^{(s)}(x)
\end{array}
\right)  =\left(
\begin{array}
[c]{cc}%
\dfrac{\Gamma(C)\Gamma(C-A-B)}{\Gamma(C-A)\Gamma(C-B)} & \dfrac{\Gamma
(C)\Gamma(A+B-C)}{\Gamma(A)\Gamma(B)}\\
\dfrac{\Gamma(2-C)\Gamma(C-A-B)}{\Gamma(1-A)\Gamma(1-B)} & \dfrac
{\Gamma(2-C)\Gamma(A+B-C)}{\Gamma(1+A-C)\Gamma(1+B-C)}%
\end{array}
\right)  \left(
\begin{array}
[c]{c}%
\mathcal{F}_{+}^{(u)}(x)\\
\mathcal{F}_{-}^{(u)}(x)
\end{array}
\right) \label{basic}%
\end{equation}

Now, taking into account also the antiholomorphic dependence and the truncated
operator product expansions (\ref{Phi12Phia}) we conclude that the
combination
\begin{equation}
C_{+}^{\text{(M)}}(\alpha_{1})C_{\text{M}}(\alpha_{1}+\beta/2,\alpha
_{2},\alpha_{3})\mathcal{F}_{+}^{(s)}(x)\mathcal{F}_{+}^{(s)}(\bar
x)+C_{-}^{\text{(M)}}(\alpha_{1})C_{\text{M}}(\alpha_{1}-\beta/2,\alpha
_{2},\alpha_{3})\mathcal{F}_{-}^{(s)}(x)\mathcal{F}_{-}^{(s)}(\bar
x)\label{inv}%
\end{equation}
is a single-valued function of $x$ if
\begin{equation}
\ \frac{C_{+}^{\text{(M)}}(\alpha_{1})C_{\text{M}}(\alpha_{1}+\beta
/2,\alpha_{2},\alpha_{3})}{C_{-}^{\text{(M)}}(\alpha_{1})C_{\text{M}}%
(\alpha_{1}-\beta/2,\alpha_{2},\alpha_{3})}=-\dfrac{\Gamma^{2}(2-C)\gamma
(C-A)\gamma(C-B)}{\gamma(1-A)\gamma(1-B)\Gamma^{2}(C)}\label{FF1}%
\end{equation}
This relation is equivalent to (\ref{F1}). The second functional relation
(\ref{F2}) is obtained through the substitution $C_{\pm}^{\text{(M)}}%
(\alpha)\rightarrow\tilde C_{\pm}^{\text{(M)}}(\alpha)$, $C_{\text{M}%
}\rightarrow\tilde C_{\text{M}}$ and $\beta\rightarrow-\beta^{-1}$.

\section{Double gamma and $\Upsilon$}

\textbf{1. Definition.} The Barnes double gamma-function $\Gamma_{2}%
(x|\omega_{1},\omega_{2})$, is defined as the analytic continuation of the
two-fold series
\begin{equation}
\log\Gamma_{2}(x|\omega_{1},\omega_{2})=\left.  \frac d{dz}\sum_{m,n=0}%
^{\infty}(x+m\omega_{1}+n\omega_{2})^{-z}\right|  _{z=0}\label{Zeta2}%
\end{equation}
(convergent at $\operatorname*{Re}z>2$) up to the point $z=0$.

\textbf{2. Contour integral. }$\Gamma_{2}(x|\omega_{1},\omega_{2})$ admits the
following integral representation
\begin{equation}
\log\Gamma_{2}(x|\omega_{1},\omega_{2})=\frac C2B_{2,2}(x|\omega_{1}%
,\omega_{2})+\frac1{2\pi i}\int_{\mathcal{C}}\frac{e^{-xt}\log(-t)}{\left(
1-e^{-\omega_{1}t}\right)  \left(  1-e^{-\omega_{2}t}\right)  }\frac
{dt}t\label{cint}%
\end{equation}
where the contour $\mathcal{C}$ goes from $+\infty$ to $+\infty$ encircling
$0$ counterclockwise, $C$ is the Euler's constant and
\begin{equation}
B_{2,2}(x|\omega_{1},\omega_{2})=\frac{(2x-\omega_{1}-\omega_{2})^{2}}%
{4\omega_{1}\omega_{2}}-\frac{\omega_{1}^{2}+\omega_{2}^{2}}{12\omega
_{1}\omega_{2}}\label{B2}%
\end{equation}
The integral is well defined if $\operatorname*{Re}\omega_{1}>0,$
$\operatorname*{Re}\omega_{2}>0$ and $\operatorname*{Re}x>0$. The function can
be analytically continued to all complex values of the periods $\omega_{1}$
and $\omega_{2}$, excluding the case $\omega_{1}/\omega_{2}$ is a real
negative number (the cases $\omega_{1}=0$ or $\omega_{2}=0$ are of course also
excluded). Otherwise (\ref{cint}) continues as a meromophic function of $x$
with no zeros and simple poles at $x=-m\omega_{1}-n\omega_{2}$, where $m$ and
$n$ are non-negative integers.

\textbf{3. Shift relations}
\begin{align}
\Gamma_{2}(x+\omega_{1}|\omega_{1},\omega_{2})  &  =\frac{\sqrt{2\pi}%
\omega_{2}^{1/2-x/\omega_{2}}}{\Gamma(x/\omega_{2})}\Gamma_{2}(x|\omega
_{1},\omega_{2})\label{G2shift}\\
\Gamma_{2}(x+\omega_{2}|\omega_{1},\omega_{2})  &  =\frac{\sqrt{2\pi}%
\omega_{1}^{1/2-x/\omega_{1}}}{\Gamma(x/\omega_{1})}\Gamma_{2}(x|\omega
_{1},\omega_{2})\nonumber
\end{align}
are readily verified.

\textbf{3. Scaling. }Function $\Gamma_{2}$ scales as follows
\begin{equation}
\Gamma_{2}(\lambda x|\lambda\omega_{1},\lambda\omega_{2})=\lambda
^{-B_{2,2}(x|\omega_{1},\omega_{2})/2}\Gamma_{2}(x|\omega_{1},\omega
_{2})\label{G2scale}%
\end{equation}
This is verified e.g., through the integral representation (\ref{cint}).

\begin{figure}
[tbh]
\begin{center}
\includegraphics[
height=4.5982in,
width=5.5383in
]%
{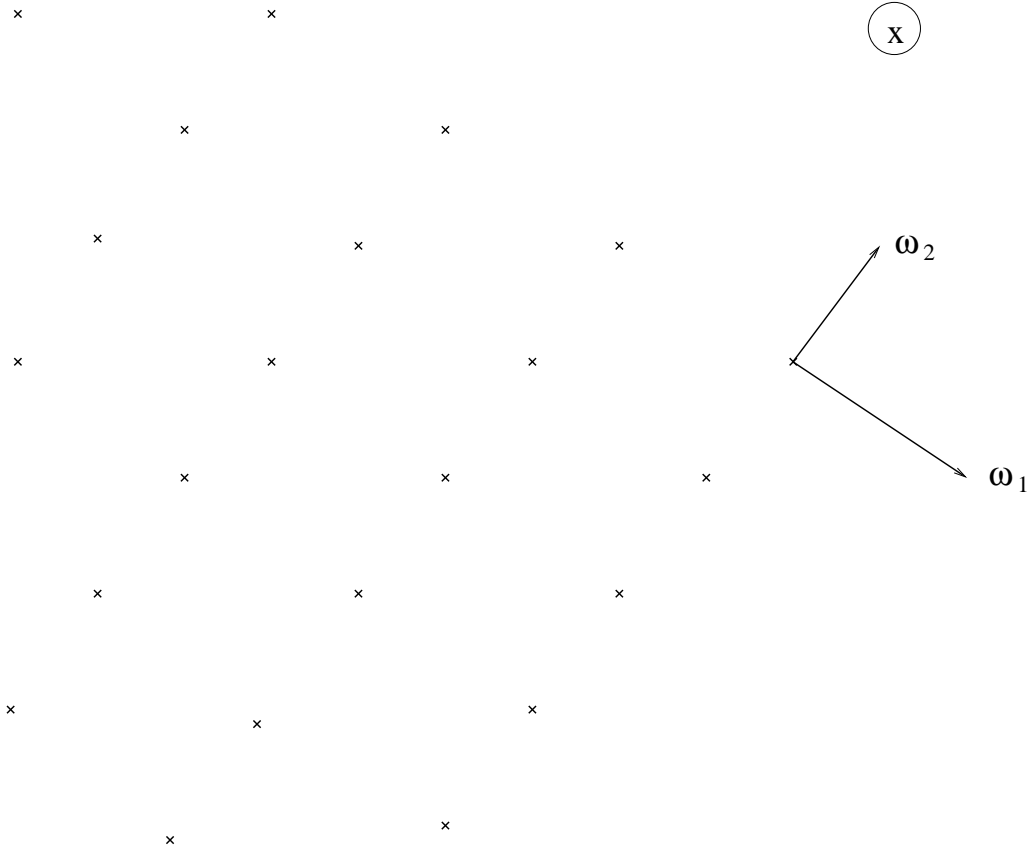}%
\caption{Poles of the double gamma function $\Gamma_{2}(x|\omega_{1}%
,\omega_{2})$.}%
\label{G2}%
\end{center}
\end{figure}

\textbf{4. Function }$\Gamma_{b}(x)$\textbf{.} In Liouville applications it's
particularly convenient to take $\omega_{1}=b$ and $\omega_{2}=b^{-1}$ and
define \cite{teschpons}
\begin{equation}
\Gamma_{b}(x)=\Gamma_{2}(x|b,b^{-1})\label{Gb}%
\end{equation}
This function is invariant under the replacement $b\rightarrow b^{-1}$
\begin{equation}
\Gamma_{b}(x)=\Gamma_{b^{-1}}(x)\label{Gbdual}%
\end{equation}
and well defined in the whole complex plane of
\begin{equation}
\tau=\omega_{2}/\omega_{1}=b^{-2}\label{tau}%
\end{equation}
except for the negative part of the real axis. For definiteness we'll always
suppose that $\operatorname*{Im}\tau\geq0$, taking advantage of (\ref{Gbdual})
otherwise. An example of the location of the poles of $\Gamma_{b}(x)$ is
plotted in fig.\ref{G2}.%

The scaling (\ref{G2scale}) always allows to express $\Gamma_{2}(x|\omega
_{1},\omega_{2})$ through $\Gamma_{b}(x)$%
\begin{equation}
\Gamma_{2}(x|\omega_{1},\omega_{2})=\left(  \omega_{1}\omega_{2}\right)
^{-B_{2,2}(x|\omega_{1},\omega_{2})/4}\Gamma_{b}((\omega_{1}\omega_{2}%
)^{-1/2}x)\label{G2Gb}%
\end{equation}
with $b=(\omega_{1}/\omega_{2})^{1/2}$.

\textbf{5. The }$\Upsilon$\textbf{-function. }$\Upsilon$ is defined through
$\Gamma_{2}$ as
\begin{equation}
\Upsilon(x|\omega_{1},\omega_{2})=\frac{\Gamma_{2}^{2}((\omega_{1}+\omega
_{2})/2|\omega_{1},\omega_{2})}{\Gamma_{2}(x|\omega_{1},\omega_{2})\Gamma
_{2}(\omega_{1}+\omega_{2}-x|\omega_{1},\omega_{2})}\label{U2}%
\end{equation}
It admits the following line integral representation
\[
\log\Upsilon(x|\omega_{1},\omega_{2})=\int_{0}^{\infty}\frac{dt}t\left[
\frac{(\omega_{1}+\omega_{2}-2x)^{2}}{4\omega_{1}\omega_{2}}e^{-2t}%
-\frac{\sinh^{2}((\omega_{1}+\omega_{2}-2x)t/2)}{\sinh(\omega_{1}%
t)\sinh(\omega_{2}t)}\right]
\]%

\begin{figure}
[tbh]
\begin{center}
\includegraphics[
height=3.096in,
width=5.7207in
]%
{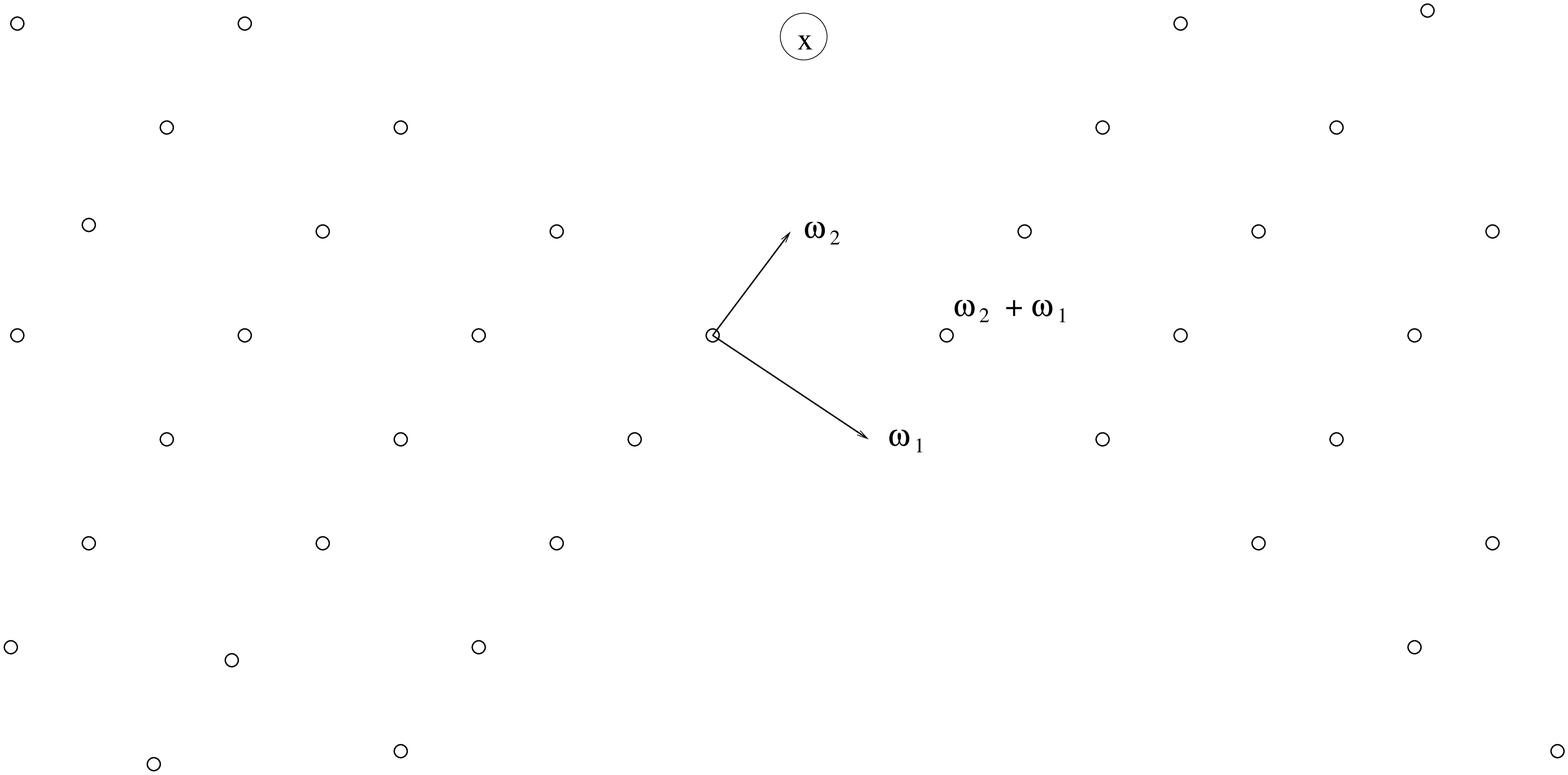}%
\caption{Position of poles of the function $\Upsilon(x|\omega_{1},\omega_{2}%
)$.}%
\label{Ypoles}%
\end{center}
\end{figure}

From (\ref{G2scale}) it follows that
\begin{equation}
\Upsilon(\lambda x|\lambda\omega_{1},\lambda\omega_{2})=\lambda^{(\omega
_{1}+\omega_{2}-2x)^{2}/(4\omega_{1}\omega_{2})}\Upsilon(x|\omega_{1}%
,\omega_{2})\label{Yscale}%
\end{equation}
while the shift relations read
\begin{align}
\Upsilon(x+\omega_{1}|\omega_{1},\omega_{2})  &  =\omega_{2}^{2x/\omega_{2}%
-1}\gamma(x/\omega_{2})\Upsilon(x|\omega_{1},\omega_{2})\label{U2shift}\\
\Upsilon(x+\omega_{2}|\omega_{1},\omega_{2})  &  =\omega_{1}^{2x/\omega_{1}%
-1}\gamma(x/\omega_{1})\Upsilon(x|\omega_{1},\omega_{2})\nonumber
\end{align}
where as usual $\gamma(x)=\Gamma(x)/\Gamma(1-x)$. Apparently $\Upsilon$
normalized in the way that
\begin{equation}
\Upsilon((\omega_{1}+\omega_{2})/2|\omega_{1},\omega_{2})=1\label{U2norm}%
\end{equation}
It's also relevant to define \cite{ZZ1}
\begin{equation}
\Upsilon_{b}(x)=\Upsilon(x|b,b^{-1})\label{Ub}%
\end{equation}
so that
\begin{equation}
\Upsilon(x|\omega_{1},\omega_{2})=(\omega_{1}\omega_{2})^{(\omega_{1}%
+\omega_{2}-2x)^{2}/(8\omega_{1}\omega_{2})}\Upsilon_{b}(\left(  \omega
_{1}\omega_{2}\right)  ^{-1/2}x)\label{U2Ub}%
\end{equation}
with $b=(\omega_{1}/\omega_{2})^{1/2}$.%

\begin{figure}
[tbh]
\begin{center}
\includegraphics[
height=2.7899in,
width=5.5408in
]%
{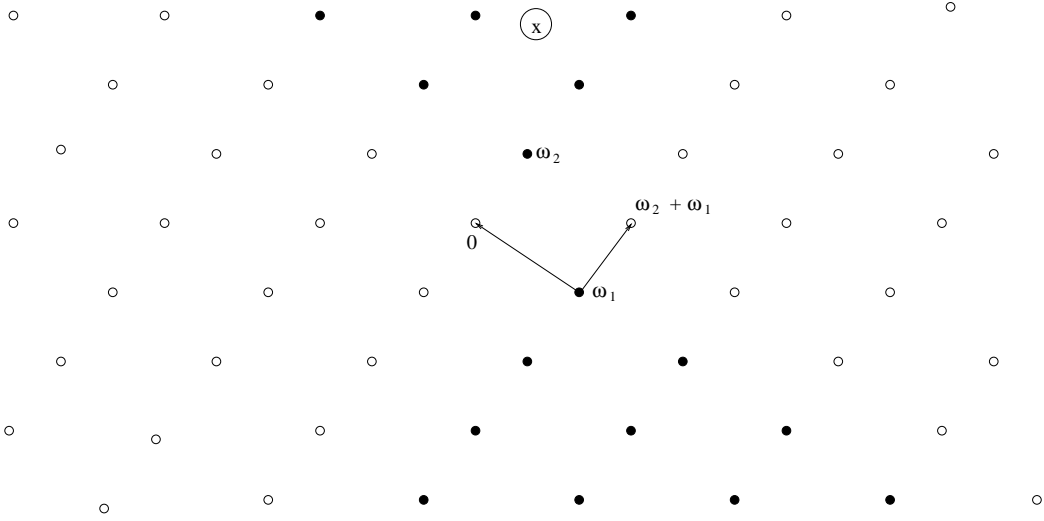}%
\caption{Zeros of the product $\Upsilon(x|\omega_{1},\omega_{2})\Upsilon
(x-\omega_{1}|e^{i\pi}\omega_{1},\omega_{2})$ in the $x $-plane. Open circles
are these of the first multiplier and filled ones are those of the second.
Together they form the regular lattice of zeros of the theta function. Arrows
show the ``periods'' $-\omega_{1}$ and $\omega_{2}$ of $\Upsilon(x-\omega
_{1}|e^{i\pi}\omega_{1},\omega_{2})$. }%
\label{theta}%
\end{center}
\end{figure}

\textbf{6. Complementarity. }Consider the following product
\begin{equation}
H(x|\omega_{1},\omega_{2})=\Upsilon(x|\omega_{1},\omega_{2})\Upsilon
(x-\omega_{1}|e^{i\pi}\omega_{1},\omega_{2})\label{H2}%
\end{equation}
Here we suppose that $\tau=\omega_{2}/\omega_{1}$ has positive imaginary part,
so that the rotation $\omega_{1}\rightarrow e^{i\pi}\omega_{1}$ goes safely
avoiding the negative real axis of $\tau.$ It is straightforward to verify
that this product is scale invariant
\begin{equation}
H(\lambda x|\lambda\omega_{1},\lambda\omega_{2})=H(x|\omega_{1},\omega
_{2})\label{Hscale}%
\end{equation}
Function $H(x|\omega_{1},\omega_{2})$ is an entire function of $x$ with the
regular lattice of zeros $x=m\omega_{1}+n\omega_{2}$, $m,n\in\mathbf{Z}$.
Together with the shift relations
\begin{align}
H(x+\omega_{1}|\omega_{1},\omega_{2})  &  =H(x|\omega_{1},\omega
_{2})\label{Hshift}\\
H(x+\omega_{2}|\omega_{1},\omega_{2})  &  =-e^{-2i\pi x/\omega_{1}}%
H(x|\omega_{1},\omega_{2})\nonumber
\end{align}
which follow from those for $\Upsilon$, this requires $H(x|\omega_{1}%
,\omega_{2})$ to have the form
\begin{equation}
H(x|\omega_{1},\omega_{2})=H_{0}e^{i\pi x/\omega_{1}}\theta_{1}\left(  \pi
x/\omega_{1}|h\right) \label{Htheta}%
\end{equation}
where $H_{0}$ is some $x$-independent constant,
\begin{equation}
h=\exp(i\pi\tau)=\exp(i\pi\omega_{2}/\omega_{1})\label{h}%
\end{equation}
and $\theta_{1}$ is the standard $\theta$-function
\begin{equation}
\theta_{1}(u|h)=i\sum_{n=-\infty}^{\infty}(-)^{n}h^{(n-1/2)^{2}}%
e^{i(2n-1)u}\label{theta1}%
\end{equation}
Normalization (\ref{U2norm}) entails $H((\omega_{1}+\omega_{2})/2|\omega
_{1},\omega_{2})=1$ and allows to determine $H_{0}$. Finally
\begin{equation}
H(x|\omega_{1},\omega_{2})=-ie^{i\pi x/\omega_{1}}\frac{\theta_{1}(\pi
x/\omega_{1}|h)}{h^{1/4}\theta_{3}(0|h)}\label{H}%
\end{equation}
where
\begin{equation}
\theta_{3}(u|h)=\sum_{n=-\infty}^{\infty}h^{n^{2}}e^{2inu}\label{theta3}%
\end{equation}

Implementing the scaling relations (\ref{Yscale}) and (\ref{Hscale}) with
$\lambda=e^{-i\pi/2}$ we arrive at
\[
H(x|b,b^{-1})=e^{-i\pi(b+b^{-1}-2x)^{2}/8}\Upsilon_{b}(x)\Upsilon
_{ib}(-ix+ib)=e^{i\pi xb^{-1}}\frac{-i\theta_{1}(\pi xb^{-1}|h)}{h^{1/4}%
\theta_{3}(0|h)}
\]
This is equivalent to relation (\ref{complementarity}) and in sect.6.

\section{``Exponential'' normalization in GMM}

Sometimes it is more convenient to use another normalization of the matter
primary fields. Let's introduce new fields $E_{\alpha}$ instead of
conventionally normalized formal GMM fields $\Phi_{\alpha}$%
\begin{equation}
\Phi_{\alpha}=\left[  \frac{\gamma(2\alpha\beta^{-1}-\beta^{-2}+2)\gamma
(\beta^{2})}{\gamma(2\alpha\beta+\beta^{2})\gamma(2-\beta^{-2})}\right]
^{1/2}\left(  \pi M\gamma(-\beta^{2})\right)  ^{-\alpha/\beta}E_{\alpha
}\label{E}%
\end{equation}
Here we also introduced an (arbitrary) ``screening coupling'' $M$ to
facilitate comparisons with the ``complex Liouville'' minded authors. Then the
special operator product expansions (\ref{Phi12Phia})
\begin{align}
E_{1,2}E_{\alpha}  &  =C_{+}^{\text{(E)}}(\alpha)\left[  E_{\alpha+\beta
/2}\right]  +C_{-}^{\text{(E)}}(\alpha)\left[  E_{\alpha-\beta/2}\right]
\label{E12Ea}\\
E_{2,1}E_{\alpha}  &  =\tilde C_{+}^{\text{(E)}}(\alpha)\left[  E_{\alpha
-\beta^{-1}/2}\right]  +\tilde C_{-}^{\text{(E)}}(\alpha)\left[
E_{\alpha+\beta^{-1}/2}\right] \nonumber
\end{align}
have the following simplified special structure constants
\begin{align}
C_{+}^{\text{(E)}}(\alpha)  &  =\tilde C_{+}^{\text{(E)}}(\alpha)=1\nonumber\\
C_{-}^{\text{(E)}}(\alpha)  &  =-\frac{\pi M}{\gamma(\beta^{2})}\frac
{\gamma(2\alpha\beta+\beta^{2}-1)}{\gamma(2\alpha\beta)}\label{CEpm}\\
\tilde C_{-}^{\text{(E)}}(\alpha)  &  =\frac{\gamma(\beta^{-2}-2\alpha
\beta^{-1}-1)}{\gamma(-2\alpha\beta^{-1})\beta^{4}(\pi M\gamma(-\beta
^{2}))^{\beta^{-2}}}\nonumber
\end{align}
We see that in this normalization the special structure constants $C_{\pm
}^{\text{(E)}}(\alpha)$ and $\tilde C_{\pm}^{\text{(E)}}(\alpha)$ are given
simply by the Dotsenko-Fateev ``screening'' integrals \cite{DF}, i.e., are
trivial if no ``screening'' is required and\footnote{The following integration
formula has been used $\int[x\bar x]^{\mu-1}[(1-x)(1-\bar x)]^{\nu-1}%
d^{2}x=\pi\gamma(\mu)\gamma(\nu)/\gamma(\mu+\nu)$.}
\begin{align}
C_{-}^{\text{(E)}}(\alpha)  &  =-M\int d^{2}x\left\langle e^{2i\alpha
\varphi(0)}e^{i\beta\varphi(1)}e^{-2i\beta\varphi(x)}e^{2i(q-\alpha
+\beta/2)\varphi(\infty)}\right\rangle _{\text{free}}\label{CEint}\\
\tilde C_{-}^{\text{(E)}}(\alpha)  &  =-\tilde M\int d^{2}x\left\langle
e^{2i\alpha\varphi(0)}e^{-i\beta^{-1}\varphi(1)}e^{2i\beta^{-1}\varphi
(x)}e^{2i(q-\alpha-\beta^{-1}/2)\varphi(\infty)}\right\rangle _{\text{free}%
}\nonumber
\end{align}
The correlation functions here are taken over the free scalar field
$\varphi(x)$ and the ``screening operators'' $e^{-2i\beta\varphi}$ and
$e^{2i\beta^{-1}\varphi}$ are weighted with the coupling constants $-M$ and
$-\tilde M$ respectively, while $\tilde M$ is determined by the following
relation
\begin{equation}
\pi\tilde M\gamma(-\beta^{-2})=\left(  \pi M\gamma(-\beta^{2})\right)
^{-\beta^{-2}}\label{Ltwiddle}%
\end{equation}
This observation makes it natural to call this normalization of the matter
fields the ``exponential'' one. We discuss now its properties, advantages and disadvantages.

\textbf{1. The two-point function} in the exponential normalization
\begin{equation}
\left\langle E_{\alpha}(x)E_{\alpha}(0)\right\rangle =\frac{G_{\alpha\alpha}%
}{\left(  x\bar x\right)  ^{2\Delta_{\alpha}}}\label{Gaa}%
\end{equation}
introduces non-trivial (albeit diagonal) metric in the space of fields with
\begin{equation}
G_{\alpha\alpha}=\frac{\gamma(2\alpha\beta+\beta^{2})\gamma(2-\beta^{-2}%
)}{\gamma(2\alpha\beta^{-1}-\beta^{-2}+2)\gamma(\beta^{2})}\left(  \pi
M\gamma(-\beta^{2})\right)  ^{2\alpha/\beta}\label{Emetric}%
\end{equation}
The identification (\ref{q-a}) now takes the form
\begin{equation}
E_{\alpha}=R_{\text{E}}(\alpha)E_{q-\alpha}\label{Eq-a}%
\end{equation}
with the following ``reflection amplitude''
\begin{align}
R_{\text{E}}(\alpha)  &  =\left(  \pi M\gamma(-\beta^{2})\right)
^{(2\alpha-q)/\beta}\frac{\gamma(2\alpha\beta+\beta^{2})\beta^{-2}}%
{\gamma(2\alpha\beta^{-1}-\beta^{-2}+2)}\label{Erefl}\\
&  =\left(  \pi M\gamma(-\beta^{2})\right)  ^{2ip/\beta}\frac{\Gamma(1+2i\beta
p)\Gamma(1-2i\beta^{-1}p)}{\Gamma(1+2i\beta^{-1}p)\Gamma(1-2i\beta
p)}\nonumber
\end{align}
In the last expression we used the ``Liouville'' parameterization
$\alpha=q+ip$.

\textbf{2. The (normalized) three-point function} now reads
\begin{align}
\  &  C_{\text{E}}(\alpha_{1},\alpha_{2},\alpha_{3})=\left\langle
E_{\alpha_{1}}E_{\alpha_{2}}E_{\alpha_{3}}\right\rangle =\left(  \pi
M\gamma(-\beta^{2})\right)  ^{\sum\alpha_{i}/\beta}\beta^{2(\beta^{-1}%
+\beta)\left(  \sum\alpha_{i}-q\right)  }\gamma(2-\beta^{-2})\gamma
(1-\beta^{2})\beta^{2}\times\nonumber\\
&  \ \ \ \frac{\Upsilon(\alpha_{1}+\alpha_{2}-\alpha_{3}+\beta)\Upsilon
(\alpha_{2}+\alpha_{3}-\alpha_{1}+\beta)\Upsilon(\alpha_{3}+\alpha_{1}%
-\alpha_{2}+\beta)\Upsilon(2\beta-\beta^{-1}+\alpha_{1}+\alpha_{2}+\alpha
_{3})}{\Upsilon(\beta)\Upsilon(2\alpha_{1}+\beta)\Upsilon(2\alpha_{2}%
+\beta)\Upsilon(2\alpha_{3}+\beta)}\label{CE}%
\end{align}

\textbf{3. No more square roots} in the exponential normalization, unlike the
case of canonical one. The general three-point function (\ref{CE}), as well as
the special structure constants (\ref{CEpm}), are meromorphic functions of the
variables $\alpha$. This seems to be an important advantage of this normalization.

\textbf{4. The structure constants} $C_{\alpha_{1},\alpha_{2}}^{\text{(E)}%
\alpha_{3}}$ in the exponential normalization are related to the corresponding
three-point function (\ref{CE}) as
\begin{equation}
C_{\alpha_{1},\alpha_{2}}^{\text{(E)}\alpha_{3}}=G^{\alpha_{3}\alpha_{3}%
}C_{\text{E}}(\alpha_{1},\alpha_{2},\alpha_{3})\label{Estruct}%
\end{equation}
Explicitly
\begin{align}
&  C_{\alpha_{1},\alpha_{2}}^{\text{(E)}\alpha_{3}}=\left(  \pi M\gamma
(-\beta^{2})\beta^{2+2\beta^{2}}\right)  ^{(\alpha_{1}+\alpha_{2}-\alpha
_{3})/\beta}\times\label{SE}\\
&  \frac{\Upsilon(\alpha_{1}+\alpha_{2}-\alpha_{3}+\beta)\Upsilon(\alpha
_{2}+\alpha_{3}-\alpha_{1}+\beta)\Upsilon(\alpha_{3}+\alpha_{1}-\alpha
_{2}+\beta)\Upsilon(2\beta-\beta^{-1}+\alpha_{1}+\alpha_{2}+\alpha_{3}%
)}{\Upsilon(\beta)\Upsilon(2\alpha_{1}+\beta)\Upsilon(2\alpha_{2}%
+\beta)\Upsilon(2\alpha_{3}+2\beta-\beta^{-1})}\nonumber
\end{align}
Comparing the analytic continuation
\begin{align}
\beta &  \rightarrow ib\label{betab2}\\
\alpha &  \rightarrow-ia\nonumber
\end{align}
of the structure constant with the corresponding structure constant in the
Liouville field theory we find
\begin{equation}
\frac{C_{-ia_{1},-ia_{2}}^{\text{(E)}-ia_{3}}(ib)}{C_{a_{1}a_{2}}%
^{\text{(L)}a_{3}}(b)}=\left(  \frac\mu M\right)  ^{(a_{1}+a_{2}-a_{3}%
)/b}T(a_{1},a_{2},a_{3})\label{Tsimple}%
\end{equation}
with double periodic $T(a_{1},a_{2},a_{3})$ from eq.(\ref{Telliptic}). Again,
the analytic relation between the GMM and Liouville characteristics is
substantially simplified in the exponential normalization. Moreover

\textbf{5. Special structure constants} $C_{\pm}^{\text{(E)}}(\alpha)$ (as
well as all other structure constants in truncated operator product
expansions) are given directly (without any normalization factors) by the
``screening integrals'' (\ref{CEint}) (or their multiple generalizations). Of
course, these integrals are analytic in all parameters and therefore the GMM
special structure constants are direct analytic continuation
\begin{align}
b  &  \rightarrow-i\beta\nonumber\\
a  &  \rightarrow i\alpha\label{bbeta}\\
\mu &  \rightarrow M\nonumber
\end{align}
of the corresponding coefficients in Liouville field theory.

\textbf{6. Singularities. }When using the exponential normalization it is
important to keep in mind that the renormalization (\ref{E}) is singular at
certain values of $\alpha$. This results in zeros and poles in the multipoint
functions, which are of coordinate nature and can be easily misinterpreted. In
addition, every primary field $\Phi_{\alpha}$ in the exponential normalization
has two representatives, $E_{\alpha}$ and $E_{q-\alpha}$, whose apparently
simple relation (\ref{Eq-a}) might also be singular. To me, this problem is
the most important disadvantage of the exponential normalization as compared
with the canonical one (\ref{CFTmn}).

\textbf{7. Gravitational two- and three-point functions} (\ref{twopointn}) and
(\ref{threepointn}) are further simplified if the exponential normalization of
the matter fields is chosen
\begin{equation}
\mathcal{U}_{a}=E_{a-b}V_{a}\label{VV}%
\end{equation}
The normalized two- and three-point functions read
\begin{align}
\frac{\left\langle \mathcal{U}_{a}\mathcal{U}_{a}\right\rangle _{\text{MG}}%
}{Z_{\text{L}}}  &  =\frac{(g-1)g(1+g)}{(1+g-2s)}L_{\text{E}}^{2}%
(a)\label{VV23}\\
\frac{\left\langle \mathcal{U}_{a_{1}}\mathcal{U}_{a_{2}}\mathcal{U}_{a_{3}%
}\right\rangle _{\text{MG}}}{Z_{\text{L}}}  &  =(g-1)g(1+g)\prod_{i=1}%
^{3}L_{\text{E}}(a_{i})\nonumber
\end{align}
where
\begin{equation}
L_{\text{E}}(a)=-\frac{\gamma(2ab-b^{2})}{M\gamma(-b^{2})}\left(  \frac{\pi
M\gamma(-b^{2})}{\pi\mu\gamma(b^{2})}\right)  ^{s}\label{legE}%
\end{equation}
, as in (\ref{gb}), we use the notations $g=b^{-2}$ and $s=ab^{-1}$.

\end{document}